\documentclass[11pt]{article}   
\usepackage[sort,move]{cite}
\usepackage{epsfig}
\usepackage{amsfonts}
\usepackage{latexsym}
\usepackage{amsmath,amssymb}
\usepackage{setspace}
\usepackage{graphicx}

\usepackage{color}
\definecolor{darkgreen}{rgb}{0,0.4,0} 
\definecolor{darkblue}{rgb}{0,0,0.4}
\definecolor{purple}{rgb}{0.4,.2,0.7}
\usepackage[colorlinks=true,citecolor=darkgreen,linkcolor=darkgreen,urlcolor=darkgreen]{hyperref} 

\usepackage[textheight=9in, textwidth=6.5in, letterpaper]{geometry}

\newcommand{\nc}{\newcommand}
\nc{\rnc}{\renewcommand}
\rnc{\bigstar}{\clubsuit}
\nc{\vq}{\vec{q}}
\nc{\bo}{\raise-0.5mm\hbox{\Large $\Box$}}
\rnc{\d}{\mathrm{d}}
\nc{\D}{\partial}
\nc{\K}{\kappa}
\nc{\bK}{\bar{\K}}
\nc{\bN}{\bar{N}}
\nc{\bq}{\bar{q}}
\nc{\bp}{\bar{p}}
\nc{\vbq}{\vec{\bar{q}}}
\nc{\g}{\gamma}
\nc{\lrarrow}{\leftrightarrow}
\nc{\rg}{\sqrt{g}}
\rnc{\[}{\begin{equation}}
\rnc{\]}{\end{equation}}
\nc{\bea}{\begin{eqnarray}}
\nc{\eea}{\end{eqnarray}}
\nc{\nn}{\nonumber}
\rnc{\(}{\left(}
\rnc{\)}{\right)}
\nc{\q}{\vec{q}}
\nc{\x}{\vec{x}}
\nc{\ep}{\epsilon}
\nc{\tto}{\rightarrow}
\rnc{\inf}{\infty}
\rnc{\Re}{\mathrm{Re}}
\rnc{\Im}{\mathrm{Im}}
\nc{\z}{\zeta}
\nc{\I}{\mathcal{I}}
\nc{\mA}{\mathcal{A}}
\nc{\A}{\mA}
\nc{\mB}{\mathcal{B}}
\nc{\mC}{\mathcal{C}}
\nc{\mD}{\mathcal{D}}
\nc{\mE}{\mathcal{E}}
\nc{\mF}{\mathcal{F}}
\rnc{\H}{\mathcal{H}}
\rnc{\L}{\mathcal{L}}
\nc{\<}{\langle}
\rnc{\>}{\rangle}
\nc{\fnl}{f_{NL}}
\nc{\fnleq}{f_{NL}^{equil.}}
\nc{\fnlloc}{f_{NL}^{local}}
\nc{\vphi}{\varphi}
\nc{\Lie}{\pounds}
\nc{\half}{\frac{1}{2}}
\nc{\bOmega}{\bar{\Omega}}
\nc{\bLambda}{\bar{\Lambda}}
\nc{\dN}{\delta N}
\nc{\gYM}{g_{\mathrm{YM}}}
\nc{\geff}{g_{\mathrm{eff}}}
\nc{\bg}{\hat{\gamma}}
\nc{\Oi}{\Omega_{[2]}}
\nc{\Oii}{\Omega_{[3]}}
\nc{\Ei}{E_{[2]}}
\nc{\Eii}{E_{[3]}}
\nc{\bOi}{\bar{\Omega}_{[2]}}
\nc{\bOii}{\bar{\Omega}_{[3]}}
\nc{\bEi}{\bar{E}_{[2]}}
\nc{\bEii}{\bar{E}_{[3]}}
\rnc{\a}{\bar{a}}
\rnc{\b}{\bar{b}}
\rnc{\c}{\bar{c}}
\rnc{\O}{\mathcal{O}}
\nc{\blambda}{\bar{\lambda}}
\nc{\oa}{\stackrel{\leftrightarrow}}
\newcommand{\lla}{\langle \! \langle}
\newcommand{\rra}{\rangle \! \rangle}

\newcommand{\p}{\partial}
\nc{\wT}{\widetilde{T}}
\rnc{\O}{\mathcal{O}}
\rnc{\L}{\mathcal{L}}
\nc{\T}{\mathcal{T}}
\nc{\Y}{\mathcal{Y}}
\nc{\X}{\mathcal{X}}
\nc{\J}{J} 

\begin{document}
\unitlength = 1mm

\begin{center}
\hfill\\
\vspace{1.5cm}
 {\LARGE {\bf Holography for inflation using conformal \\[1ex] perturbation theory}} 

\vspace{1cm}

{\large Adam Bzowski,$^{a,c}$ Paul McFadden$^b$ and Kostas Skenderis$^{a,c,d}$}

\vspace{0.7cm}   

{\small
$^a${\it School of Mathematics, University of Southampton, UK.} \\
$^b${\it Perimeter Institute for Theoretical Physics, Waterloo, Canada.}\\
$^c${\it Korteweg--de$\,$Vries Institute for Mathematics,}
$^d${\it Institute for Theoretical Physics, Amsterdam, Netherlands.}\\
}

\vspace{0.5cm}
{\small E-mail: {\tt a.w.bzowski@uva.nl, pmcfadden@perimeterinstitute.ca, k.skenderis@soton.ac.uk} }

\vspace{1.0cm}
\end{center}

\begin{center}
{\bf Abstract}
\end{center}

We provide a precise and quantitative holographic description of a class of inflationary slow-roll models.
The dual QFT is a deformation of a three-dimensional CFT by a nearly marginal operator, which, in the models we consider, generates an RG flow to a nearby IR fixed point.
These models describe hilltop inflation, where the inflaton rolls from a local maximum of the potential in the infinite past (corresponding to the IR fixed point of the dual QFT) to reach a nearby local minimum in the infinite future (corresponding to the UV of the dual QFT).
Through purely holographic means, we compute the spectra and bispectra of scalar and tensor cosmological perturbations. The QFT correlators to which these observables map holographically may be 
calculated using conformal perturbation theory, even when the dual QFT is strongly coupled. 
Both the spectra and the bispectra may be expressed this way in terms of CFT correlators that are fixed, up to a few constants, by conformal invariance.  The form of slow-roll inflationary correlators is thus determined by the perturbative breaking of the de Sitter isometries away from the fixed point.  Setting the constants to their values obtained by AdS/CFT at the fixed point,
we find exact agreement with known expressions for the slow-roll power spectra and non-Gaussianities.

\pagestyle{empty}

\pagebreak 
\setcounter{page}{1}
\pagestyle{plain}
\setcounter{tocdepth}{2}

\hrule

\tableofcontents

\bigskip

\vspace{0.1cm}

\hrule

\bigskip

\section{Introduction}

The primordial perturbations encode a great wealth of information about the early universe, and, as such, 
it is important to understand their structure as far as possible. In particular, it is important to understand which features 
are fixed by symmetries and which are a property of the specific fundamental theory that governs the universe at early times.
Since the spacetime geometry during slow-roll inflation is quasi-de Sitter, one may anticipate that at least some of the properties of 
cosmological observables are fixed by the underlying broken de Sitter isometries, and indeed some of the recent literature 
is devoted to answering this question \cite{Cheung:2007st,Antoniadis:2011ib,Maldacena:2011nz,Hinterbichler:2011qk,Creminelli:2011mw,Hinterbichler:2012mv,Creminelli:2012ed,
Hinterbichler:2012nm,Assassi:2012zq,Kehagias:2012pd,Kehagias:2012td,Assassi:2012et}. In this paper we use a holographic set up to address this question, and we find that essentially all spectra and bispectra, for both scalars and tensors, are fixed by the broken de Sitter isometries up to a number of constants. While our results are obtained within a class of models, we believe the answer holds more generally.

We will use the holographic framework we developed in our earlier work \cite{McFadden:2009fg, McFadden:2010na, McFadden:2010jw, McFadden:2010vh,McFadden:2011kk}. This framework is applicable when the spacetime 
is either asymptotically power-law or asymptotically de-Sitter.  In previous works we presented a complete analysis of the power-law
models (in particular their phenomenology \cite{Easther:2011wh, Bzowski:2011ab}, see also \cite{Dias:2011in, Coriano:2012hd}) and 
here we will focus exclusively on the asymptotically de Sitter case. 
Earlier studies of this case include \cite{Strominger:2001gp, Larsen:2002et, Halyo:2002zg, Larsen:2003pf, vanderSchaar:2003sz, Larsen:2004kf, Seery:2006tq}
building on the dS/CFT correspondence \cite{Strominger:2001pn,Witten:2001kn,Maldacena:2002vr}\footnote{See 
\cite{Harlow:2011ke, Dong:2011uf, Anninos:2011ui,Anninos:2011af,Hertog:2011ky, Hartle:2012qb,Anninos:2012ft,Hartle:2012tv, Castro:2012gc, Marolf:2012kh} for a sample of more recent works.}.

In our previous work,  we proved 
that one can express all spectra and bispectra of inflationary models in terms 
of the correlation functions of a dual three-dimensional QFT, after appropriate analytic continuation in parameters and momenta.
When gravity is weakly coupled, so that standard inflationary computations are valid, the dual QFT is at strong coupling.
The proof in our earlier work assumed the validity of standard AdS/CFT duality which is used in order to compute the QFT 
correlators at strong coupling. Here we will relax this assumption and directly compute these correlators on the QFT side
using conformal perturbation theory \cite{Ludwig:1987gs,Zamolodchikov:1987ti}. As such our results are valid both when the QFT is at strong coupling and at weak coupling, with the latter corresponding to a universe that was non-geometric at early times.

Since the group of  de Sitter isometries is the same as the (Euclidean) conformal group in one dimension less, an asymptotically de Sitter spacetime will have a QFT dual that in the UV becomes conformal. This QFT may either be a deformation of a  CFT, or else 
a CFT in a non-trivial state that spontaneously breaks conformal invariance. It appears that the  spacetimes
corresponding to the second option cannot satisfy the slow-roll conditions, so in this paper we will focus on the first option.
Once the CFT is deformed by a relevant operator it undergoes an RG flow, and the inverse of this evolution corresponds to 
cosmic evolution in the bulk.  
Of the various possible fates for the RG flow, we focus here on the case where the flow leads us to a new fixed point in the IR.
The RG flow in the vicinity of this IR fixed point then gives rise to a red-tilted cosmological power spectrum in line with current observational preferences based on a minimal power-law $\Lambda$CDM fit \cite{Komatsu:2010fb}.  Our models will therefore have an asymptotically de Sitter epoch in the far past as well as in the far future.  

In the vicinity of either fixed point, the RG flow will be controlled by the most nearly marginal operator in the theory; about the UV fixed point this operator will be marginally relevant, while about the IR it will be marginally irrelevant.  
As we wish to construct a single-field model of inflation, we will assume this most nearly marginal operator to be a single scalar.  For simplicity, we will further take this single scalar operator to control the entire RG flow from the UV to the IR fixed point.  
A more sophisticated holographic model incorporating the effects of reheating followed by epochs of radiation and matter domination would presumably relax this assumption, allowing other operators to enter and eventually to dominate the RG flow terminating the period of single-field inflation.  
Nevertheless, the simple model we study here will capture the same long-wavelength behaviour found in a more complete model.

To be able to recover standard inflationary physics from this
scenario, it is necessary to be able to perform explicit calculations in
the dual QFT.  The strongly coupled nature of this QFT at first sight
amounts to an insurmountable obstacle.
Our starting point in this paper is the observation that, for an RG
flow connecting two {\it nearby} fixed points, 
it is possible to perform perturbation theory in the small parameter controlling the separation of these fixed points.\footnote{Well-known examples of this include the $1/N$ and $\ep$-expansions for the RG flow between the Gaussian and Wilson-Fisher fixed points in dimensions less than four.} A priori, this small parameter has nothing 
to do with the coupling constant of the QFT, which in this case will be large.  One may thus perturbatively compute correlators in terms of the CFT correlators associated with either of the fixed points.  As the 2- and 3-point correlators of a CFT are universal, depending only on the dimension and OPE coefficients of the operators involved, the 2- and 3-point correlators of the QFT are also fully fixed in terms of these quantities.   Ultimately, the dimension and OPE coefficient of the scalar operator dual to the inflaton will map to the slow-roll parameters of the dual cosmology. 

The observed small deviation of the power spectrum from exact scale invariance 
 plays a double role in these models: on the one hand it controls the dimension of the nearly marginal scalar operator dual to the inflaton,
while on the other hand, it also controls the separation of the UV and IR fixed points in the space of couplings.
More precisely, if the UV dimension of the deforming operator is $\Delta_{UV}=3-\lambda$, where $0<\lambda\ll 1$, the IR dimension at leading order is $\Delta_{IR}=3+\lambda$, while the separation in coupling between the fixed points is proportional to $\lambda$.  Cosmologically, on long wavelengths the tilt of the power spectrum is governed by the IR dimension and we obtain a red tilt $n_S-1=-2\lambda$, while short wavelengths probe the UV fixed point yielding a blue tilt $n_S-1=+2\lambda$.  Since in more complete models the fate of the RG flow in the UV may be different, it is the red-tilted long-wavelength portion that is of principal interest.

Computing the 2- and 3-point correlators in the dual QFT to leading order in $\lambda$ and inserting these in the holographic formulae derived in \cite{McFadden:2009fg, McFadden:2010vh, McFadden:2011kk}, we recover scalar and tensor power spectra and non-Gaussianities of exactly the form generated by a period of slow-roll inflation.\footnote{The slow-roll 3-point function for three gravitons is insensitive to the scalar deformation we study here, however, and as such may be derived from a dual QFT which is an exact CFT \cite{Maldacena:2011nz,Bzowski:2011ab}.  As the treatment in these two papers is already complete we will omit this correlator from our present study.}  
In fact, it is straightforward to systematically identify the relevant class of inflationary potentials as those deriving from a cubic polynomial superpotential.  
Many simple generalisations of this model are possible, both through the action of field redefinitions and through the exploration of related potentials.

We thus  have a class of backgrounds for which it is possible to directly compute on both sides of the holographic correspondence.
The exact agreement we find between conventional slow-roll correlators calculated through the gravity description and those calculated holographically through the dual QFT constitutes a highly non-trivial test of holographic cosmology.

The outline of this paper is as follows.  Section \ref{sec:conf_pert_th} is devoted to conformal perturbation theory: we present an introduction to perturbative RG flows and explain in detail the computation of 2- and 3-point correlation functions.  
In Section \ref{sec:hol_calc}, we summarise briefly the framework for holographic cosmology proposed in \cite{McFadden:2009fg}, then present the necessary holographic formulae for the cosmological power spectra and non-Gaussianities in terms of the QFT correlators computed by conformal perturbation theory.  Applying these holographic formulae, we arrive at our holographic predictions for the cosmological observables.  In Section \ref{sec:Ident}, we identify the bulk inflationary action and solve for the background evolution and slow-roll parameters, confirming the slow-roll cosmological correlators obtained holographically.
We conclude with a discussion in Section \ref{sec:disc}.  Two appendices contain additional technical details concerning the Fourier transform of CFT correlators from position space to momentum space, and a list of relevant QFT Ward identities.

{\it Note added:} While this paper was being finalised, \cite{Schalm:2012pi} appeared which contains some overlap with Section \ref{sec:3scalarshol}.

\section{Conformal perturbation theory}
\label{sec:conf_pert_th}

\subsection{Perturbative RG flows}

We consider a three-dimensional Euclidean conformal field theory perturbed by the addition of a nearly marginal scalar operator of dimension $\Delta = 3-\lambda$, where $0<\lambda \ll 1$.
The action takes the form
\[\label{Sdef}
S = S_{CFT}+\int d^3 x \sqrt{g} \vphi\Lambda^{-\lambda} \O,
\]
where $\vphi(x)$ represents the dimensionless bare coupling at some energy scale $1/\Lambda$ close to the UV fixed point, and $\O(x)$ is the bare scalar operator of dimension $\Delta$.
We have introduced a non-trivial background metric $g_{ij}(x)$ acting as a source for the stress tensor $T_{ij}$.
After differentiating with respect to $g_{ij}$ a sufficient number of times to bring down the appropriate insertions of $T_{ij}$ we will then set $g_{ij}=\delta_{ij}$.
With this accomplished, correlators in the perturbed theory differ from those in the unperturbed CFT by an insertion of 
\begin{align}
\exp\Big[-\int\d^3x \vphi \Lambda^{-\lambda} \O \Big] 
& = \sum_{n=0}^\infty \frac{1}{n!}(-\vphi \Lambda^{-\lambda})^n \int d^3 x_1\ldots \d^3x_n\, \O(x_1)\ldots\O(x_n), 
\end{align}
where the range of integration must be cut off so that no two operator insertions are closer than $\Lambda$.
Remarkably, we will find that all terms in this sum contribute at leading order meaning the entire series must be resummed.

On a flat background, the unperturbed CFT correlators take the form
\begin{align} \label{CFT_2n3pt_fns}
\<\O(x_1)\O(x_2)\>_0 = \frac{\alpha}{|x_{12}|^{2\Delta}}, \quad
\<\O(x_1)\O(x_2)\O(x_3)\>_0 &= \frac{\alpha\, C}{|x_{12}|^{\Delta} |x_{23}|^{\Delta} |x_{31}|^{\Delta}}, 
\end{align}
where $x_{ij} \equiv x_i-x_j$ and we use the subscript zero to distinguish correlators in the unperturbed CFT from correlators in the perturbed theory.
The constant $\alpha$ encodes the arbitrary normalisation of the 2-point function, while the constant $C$ appears in the OPE
\begin{align}\label{OO_OPE}
\O(x_1)\O(x_2) &= \frac{\alpha}{|x_{12}|^{6-2\lambda}} + \frac{C}{|x_{12}|^{3-\lambda}}\,\O(x_2)  + \ldots \quad \mathrm{as}\,\,\,\,|x_{12}|\tto 0. 
\end{align}
\begin{figure}[t]
\center
\includegraphics[width=8cm]{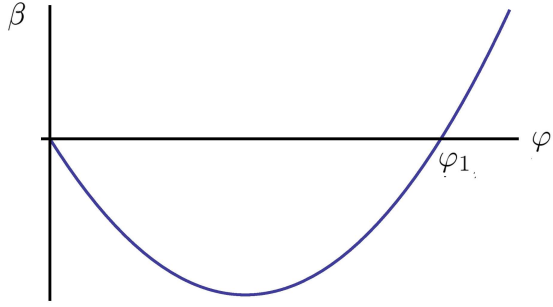} 
\caption{\label{betaplot} When the OPE coefficient $C$ is a positive constant of order unity the $\beta$-function has a UV fixed point at the origin and a nearby IR fixed point at $\vphi=\vphi_1\ll 1$.}
\end{figure}

The $\beta$-function for the coupling $\vphi$ may be found by requiring invariance of the partition function under changes of $\Lambda$ (see, e.g., \cite{Polyakov:1987ez}, or the more recent \cite{Klebanov:2011gs}),   
yielding
\[\label{beta_fn}
\beta \equiv -\frac{\d\vphi}{\d \ln \Lambda} = -\lambda \vphi + 2\pi C \vphi^2 +O(\vphi^3).
\]
Assuming $C$ to be a positive constant of order unity, the $\beta$-function is as illustrated in Fig.~\ref{betaplot} and we obtain an RG flow to a nearby IR fixed point at
\[\label{vphi1_def}
\vphi = \vphi_1 +O(\lambda^2), \qquad \vphi_1\equiv \frac{\lambda}{2\pi C} \ll 1.
\]
If instead $C$ vanishes or is negative, then the nature of the IR theory will depend on the higher order coefficients in the $\beta$-function; we will not consider these cases here.\footnote{\label{betan} Nevertheless, we anticipate a simple generalisation to cases where  $\beta = -\lambda\vphi+ C_n \vphi^n+O(\vphi^{n+1})$, cf.~footnote \ref{Wn}.} 
For positive $C$ then, since $\vphi$ is small throughout the flow, we may remove higher order terms in the $\beta$-function by a suitable field redefinition $\vphi =\bar{\vphi}+O(\bar{\vphi}^3)$.  In the following, we will assume this has been accomplished and work with the purely quadratic $\beta$-function.  Results for the general case may then be found by undoing the field redefinition, generating corrections at subleading orders in the expansion parameter $\lambda$.

Expanding the quadratic $\beta$-function about the IR fixed point, we find
\[\label{betaIR}
\beta = \lambda(\vphi-\vphi_1) + 2\pi C (\vphi-\vphi_1)^2.
\]
In the IR CFT, $\O$ thus has dimension $\Delta_{IR}=3+\lambda$ while the OPE coefficient is unchanged.

The entire RG flow may be obtained by integrating the $\beta$ function directly, yielding
\[\label{RGflow}
\Big(\frac{\Lambda_0}{\Lambda}\Big)^\lambda =\frac{\vphi_0}{\vphi}\frac{(\vphi_1-\vphi)}{(\vphi_1-\vphi_0)},
\]
where we imposed the boundary condition $\vphi(\Lambda_0) =\vphi_0$ for $0< \vphi_0 <\vphi_1$.
Inverting, we find
\[\label{phibg}
\vphi = \vphi_1 \Big[1+\frac{\vphi_1}{\phi} \Lambda^{-\lambda}\Big]^{-1},
\]
where 
\[\label{renphi}
\phi = \vphi_1 \Lambda_0^{-\lambda} \Big[\frac{\vphi_1}{\vphi_0}-1\Big]^{-1}\hspace{-1mm}.
\]
Consequently, as we remove the cutoff,  
\[
\vphi \tto \phi \Lambda^\lambda \qquad \mathrm{as}\qquad \Lambda \tto 0,
\]
allowing us to identify $\phi$ as the dimensionful renormalised coupling in the UV CFT.

\subsection{Scalar 2-point function}
\label{sec:scalar2pt}

Let us now compute the 2-point function of $\O$ in the perturbed theory,
\[\label{OO_Sum}
\<\O(x_1)\O(x_2)\> = \sum_{n=0}^{\infty}\frac{1}{n!}(-\vphi \Lambda^{-\lambda})^n\I_n, 
\]
where $\I_n$ is an unperturbed CFT correlator with $n$ integrated scalar insertions,
\[\label{In_def}
\I_n = \int\d^3z_1\ldots\d^3z_n \<\O(x_1)\O(x_2)\O(z_1)\ldots\O(z_n)\>_0.
\]
To regulate the integral the range of integration is restricted so that no two insertion points approach closer than the cutoff distance $\Lambda$.
As our intention is to work to leading order in $\lambda$, it is sufficient to compute only the leading singular behaviour of $\I_n$ as $\lambda \tto 0$.  We will see by the argument to follow that $\I_n \sim \lambda^{-n}$ in this limit; combined with the prefactor ${\vphi}^n \sim \lambda^n$, each term in the sum \eqref{OO_Sum} then makes an order one contribution.

Beginning with the integral $\I_1$,  
we may formally impose the cutoff by inserting two Heaviside step functions,
\[\label{I1def}
\I_1 = \int \d^3 z_1 \<\O(x_1)\O(x_2)\O(z_1)\>_0 \Theta(|z_1-x_1|-\Lambda)\Theta(|z_1-x_2|-\Lambda).
\]
As we are not interested in contact terms in the 2-point function \eqref{OO_Sum} we will assume that $|x_{12}| \gg \Lambda$.  
If we now vary with respect to the cutoff $\Lambda$, we pick up contributions from the two spherical shells surrounding $x_1$ and $x_2$,
\begin{align} \label{I1_ODE}
\frac{\d \I_1}{\d \Lambda} &= \int \d^3 z_1 \< \O(x_1)\O(x_2)\O(z_1)\>_0 [-\delta(|z_1-x_1|-\Lambda)-\delta(|z_1-x_2|-\Lambda)]\nn\\
& = - 2 (4\pi \Lambda^2) \frac{C}{\Lambda^{3-\lambda}} \<\O(x_1)\O(x_2)\>_0 + \ldots,
\end{align}
where in the second line we used the OPE \eqref{OO_OPE}.  
As $\<\O(x_1)\O(x_2)\>_0$ is independent of $\Lambda$, upon integrating we obtain
\[\label{I1eq}
\I_1 = -\frac{8\pi C}{\lambda}(\Lambda^\lambda-f |x_{12}|^\lambda) \<\O(x_1)\O(x_2)\>_0   + \ldots,
\]
where $f$ is an arbitrary constant, the $|x_{12}|^\lambda$ dependence being fixed on dimensional grounds as no other scales are present.
The ellipsis indicates omitted contributions from the remaining terms in the OPE \eqref{OO_OPE}.  Crucially, these contributions cannot take the form of $1/\lambda$ poles unless there are other terms in the OPE scaling as $|x_{12}|^{-3+m\lambda}$ for some nonzero constant $m$.  As a simplifying assumption, we will therefore assume that such terms, if present at all, are of subleading order in $\lambda$, i.e., the associated OPE coefficient is of order $\lambda$ or greater.\footnote{An exception to this will be the stress tensor, although as we will see in Section \ref{sec:Tintro} its inclusion does not affect our present results.}
Physically, this means that at leading order $\O$ is the only operator becoming marginal in the limit $\lambda\tto 0$.
The equation \eqref{I1eq} then captures the leading behaviour in this limit. 

To determine the constant of integration $f$ we require this limit to be non-singular, fixing $f=1$.
As $\lambda\tto 0$ we then obtain a logarithmic dependence on the cutoff $\Lambda$ signalling a Weyl anomaly,
\[
\lim_{\lambda\tto 0}\I_1 = 8\pi C \ln(|x_{12}|/\Lambda) \<\O(x_1)\O(x_2)\>_0 + \ldots.
\]
For $\lambda>0$, on the other hand, there is no Weyl anomaly and we may safely remove the cutoff. Sending $\Lambda\tto 0$, we find
\[\label{I1_2}
\I_1 = \frac{8\pi C}{\lambda} \frac{\alpha}{|x_{12}|^{6-3\lambda}}+\ldots 
\]
The apparently singular behaviour of this result as $\lambda\tto 0$ is simply an artefact of removing the cutoff.

Proceeding now to the general integral $\I_n$, we first introduce step functions to regulate the separation between all possible pairs of insertion points enforcing $|z_{ij}|>\Lambda$, $|z_i-x_1|>\Lambda$ and $|z_i-x_2|>\Lambda$ for all $i,j=1\ldots n$ such that $i<j$.  Differentiating with respect to the cutoff, in place of \eqref{I1_ODE} we now obtain
\[\label{In_ODE}
\frac{\d \I_n}{\d \Lambda} = - 4\pi \Lambda^2 B_n \frac{C}{\Lambda^{3-\lambda}} \I_{n-1} + \ldots,
\]
where the combinatorial factor $B_n$ counts the number of step functions we had initially, i.e., the number of pairs we can form by bringing together the $n$ insertion points $z_i$, either amongst themselves or with either $x_1$ or $x_2$, namely
\[\label{Bn}
B_n = \binom{n}{2} + 2n = \frac{1}{2}n(n+3).
\]
Note that in writing \eqref{In_ODE} we have effected a dilute gas approximation in which contributions to the integral $\I_n$ from configurations in which more than two insertion points coincide are neglected.  
This approximation is justified since the phase space associated with these configurations is comparatively small while the value of the integrand is comparable.

Prior to integrating \eqref{In_ODE}, it is useful to trade $\Lambda$ for
\[
y = 1- \Big(\frac{\Lambda}{|x_{12}|}\Big)^\lambda,
\]
so that 
\[
\frac{\d \I_n}{\d y} = n(n+3)\frac{2\pi C}{\lambda} |x_{12}|^\lambda \I_{n-1} + \ldots
\]
To fix the arbitrary constant of integration, we then require that $\lambda^n \I_n \tto 0$ as $\lambda \tto 0$, i.e., the constant is chosen so as to cancel the leading $1/\lambda^{n}$ pole as $\lambda\tto 0$.  Divergences due to subleading poles will be cancelled by the subleading terms we omitted in \eqref{In_ODE}.
Given that $y\tto 0$ as $\lambda \tto 0$, we find that at each order $n$ the constants arising from integration with respect to $y$ vanish, yielding 
\[\label{resultforIn}
\I_n = \frac{n! (n+3)!}{3!} \Big(\frac{2\pi C}{\lambda} |x_{12}|^\lambda\Big)^n  \frac{y^n}{n!}\I_0 +\ldots 
= \frac{\alpha}{6} (n+3)! \,\vphi_1^{-n} y^n |x_{12}|^{(n+2)\lambda-6} +\ldots,
\]
with $\vphi_1$ as given in \eqref{vphi1_def}.

The 2-point function in the perturbed theory may now be evaluated at leading order in $\lambda$ courtesy of \eqref{OO_Sum},
\begin{align}
\<\O(x_1)\O(x_2)\> &= \frac{\alpha}{6}\, |x_{12}|^{2\lambda-6}\, \sum_{n=0}^\infty (n+3)(n+2)(n+1) \Big[{-}\frac{\vphi}{\vphi_1}\Big(\frac{|x_{12}|^\lambda}{\Lambda^\lambda}-1\Big)\Big]^n. 
\end{align}
Sending $\Lambda\tto 0$ (taking note of the $\Lambda$-dependence of $\vphi$ in \eqref{phibg}), we may re-express this result as a sum of exact CFT 2-point functions with shifted dimensions:
\[\label{OObin}
\<\O(x_1)\O(x_2)\> = \frac{1}{6}\sum_{n=0}^\infty(n+3)(n+2)(n+1)\Big({-}\frac{\phi}{\vphi_1}\Big)^n \<\O_{\Delta_n}(x_1)\O_{\Delta_n}(x_2)\>_0,
\]
where $\phi$ is the renormalised coupling defined in \eqref{renphi} and
\[
\Delta_n=\Delta-\frac{n\lambda}{2} = 3-\frac{\lambda}{2}(n+2).  
\]
Summing up the binomial series, we find
\[\label{2pt_result}
\<\O(x_1)\O(x_2)\>=
\alpha |x_{12}|^{2\lambda-6}\,\Big[1+\frac{\phi}{\vphi_1}|x_{12}|^\lambda\Big]^{-4}.
\]
Finally, to transform to momentum space, starting from \eqref{OObin} we use the result
\[
\<\!\<\O_{\Delta_n}(q)\O_{\Delta_n}(-q)\>\!\>_0 = 
\int \d^3 x_{12} \, |x_{12}|^{-6+(n+2)\lambda} e^{-iq\cdot x_{12}} = \frac{\pi^2}{12}\, q^{3-(n+2)\lambda}(1+O(\lambda)),
\]
and then resum to find
\[\label{OO_FT}
\<\!\<\O(q)\O(-q)\>\!\> = \frac{\pi^2}{12}\, \alpha q^{3-2\lambda} \Big[1+\frac{\phi}{\vphi_1}q^{-\lambda}\Big]^{-4}.
\]
Here, and throughout the paper, our notation $\lla \ldots \rra$ indicates momentum space correlators in which the overall delta function $(2\pi)^3\delta(\sum_iq_i)$ from momentum conservation has been removed.
The result \eqref{OO_FT} is correct to leading order in $\lambda$ after expansion in the renormalised coupling $\phi$.

\subsection{Scalar 3-point function}

In the case where all the $|x_{ij}|^\lambda$ are of comparable magnitude, i.e., when there is effectively a single scale $L^\lambda = |x_{12}|^\lambda(1+O(\lambda)) = |x_{23}|^\lambda(1 +O(\lambda)) = |x_{31}|^\lambda(1+O(\lambda))$, our arguments above may be straightforwardly generalised to yield the leading order 3-point function for separated insertions,
\[\label{3OL}
\<\O(x_1)\O(x_2)\O(x_3)\> = \frac{\alpha\,C}{|x_{12}|^\Delta |x_{23}|^\Delta |x_{31}|^\Delta}\,\Big[1+\frac{\phi}{\vphi_1}L^\lambda\Big]^{-6}.
\]
The power of minus six appearing in this result arises because here we are summing a binomial series derived from the combinatorial factor $B_n=\binom{n}{2}+3n$ in place of \eqref{Bn}, encoding the presence of an additional fixed scalar insertion.
Note also that, since the dimension of $\O$ differs from the spatial dimension only by $\lambda$, the general CFT correlator with an arbitrary number of integrated scalar insertions is invariant under special conformal transformations at leading order in $\lambda$, constraining the arbitrary functions arising from integrating with respect to the cutoff to be of the form $L^\lambda$.

If instead the $|x_{ij}|^\lambda$ are no longer all of comparable magnitude, we find ourselves in a limit where the OPE is applicable, and utilising our previous result \eqref{2pt_result}
we find, e.g., when $|x_{12}|^\lambda\ll |x_{23}|^\lambda\approx |x_{31}|^\lambda$,  
\[\label{sqOOOx}
\<\O(x_1)\O(x_2)\O(x_3)\> = \frac{\alpha\,C}{|x_{12}|^\Delta |x_{23}|^{2\Delta}}\,\Big[1+\frac{\phi}{\vphi_1}|x_{23}|^\lambda\Big]^{-4}.
\]
We may then combine \eqref{3OL} and the three limiting cases of the form \eqref{sqOOOx} into a single result applicable at leading order for all configurations,
\[\label{3pt_pos}
\<\O(x_1)\O(x_2)\O(x_3)\> =  \alpha \,C \prod_{i<j} |x_{ij}|^{-\Delta}\Big[1+\frac{\phi}{\vphi_1}|x_{ij}|^\lambda\Big]^{-2}.
\]
Expanding out the binomial series, the leading order 3-point function in the perturbed theory may alternatively be expressed as a sum of exact CFT 3-point functions with shifted dimensions,
\[\label{OOO_as_sum}
\<\O(x_1)\O(x_2)\O(x_3)\> = \sum_{\ell_1,\ell_2,\ell_3=0}^\infty(\ell_1+1)(\ell_2+1)(\ell_3+1)\Big({-}\frac{\phi}{\vphi_1}\Big)^{\ell_t}
\<\O_{\Delta_1}(x_1)\O_{\Delta_2}(x_2)\O_{\Delta_3}(x_3)\>_0,
\]
where
\[
\Delta_i = \Delta - \frac{\lambda}{2}(\ell_t-\ell_i), \qquad \ell_t = \ell_1+\ell_2+\ell_3.
\]
This expanded form of the 3-point function is useful for performing the Fourier transform to momentum space,
as we must do in order to ultimately connect with standard inflationary results.
We discuss the details of this Fourier transform in Appendix \ref{app:OOO_FT}.  
At leading order in $\lambda$, the result is
\begin{align}\label{OOO_FT}
\<\!\<\O(q_1)\O(q_2)\O(q_3)\>\!\> &= \frac{\pi^3 \alpha\,C}{3\lambda} q_3^{-\lambda}\big[1+\frac{\phi}{\vphi_1}q_3^{-\lambda}\big]^{-1} 
\Big(\sum_{j=1}^3 q_j^{3-2\lambda}\Big[1+\frac{\phi}{\vphi_1} q_j^{-\lambda}\Big]^{-4}\Big)
\nn\\ &
=\frac{4\pi C}{\lambda}  q_3^{-\lambda}\Big[1+\frac{\phi}{\vphi_1}q_3^{-\lambda}\Big]^{-1} \sum_{j=1}^3 \<\!\<\O(q_j)\O(-q_j)\>\!\>,
\end{align}
where the reference momentum $q_3$ is chosen as the largest of the three momenta and hence is implicitly nonzero (see Appendix \ref{app:OOO_FT} for details).

As a check on our calculations, note that in the squeezed limit where we take one of the remaining momenta to zero, we obtain
\[\label{sq1}
\<\!\<\O(0)\O(q)\O(-q)\>\!\> = \frac{8\pi C}{\lambda}  q^{-\lambda}\Big[1+\frac{\phi}{\vphi_1}q^{-\lambda}\Big]^{-1} \<\!\<\O(q)\O(-q)\>\!\>.
\]
This result is of precisely the expected form, since 
\begin{align}
-\frac{\p}{\p\phi}\<\O(x_1)\O(x_2)\> &= -\frac{\p}{\p\phi}\<\O(x_1)\O(x_2)e^{-{\int}\phi\O}\>_0 = \int \d^3 z \<\O(x_1)\O(x_2)\O(z) e^{-{\int} \phi \O}\>_0 \nn \\ &=
\int \d^3z\, \<\O(x_1)\O(x_2)\O(z)\>,
\end{align}
and hence the zero-momentum limit
\begin{align}\label{zml}
\<\!\<\O(0)\O(q)\O(-q)\>\!\> = -\frac{\p}{\p\phi}\<\!\<\O(q)\O(-q)\>\!\>. 
\end{align}
Inserting our earlier result \eqref{OO_FT} for the 2-point function in momentum space, we recover precisely \eqref{sq1}.

In fact, on dimensional grounds $\lla \O(q)\O(-q)\rra = q^{3-2\lambda}F(\phi q^{-\lambda})$ for some function $F$, since when $\phi$ vanishes the 2-point correlator in the perturbed theory must reduce to that of the exact CFT, for which $\lla \O(q)\O(-q)\rra_0 \sim q^{3-2\lambda}$.  This yields the Callan-Symanzik equation
\[
0=\Big(\frac{\p}{\p \ln q}+\lambda\phi\frac{\p}{\p\phi}-3+2\lambda\Big)\lla\O(q)\O(-q)\rra,
\]
which, when combined with \eqref{zml}, gives 
\[\label{sqformOOO}
\lambda\phi \<\!\<\O(0)\O(q)\O(-q)\>\!\> = \Big(-3+2\lambda+\frac{\p}{\p \ln q}\Big) \lla \O(q)\O(-q)\rra .
\]
This result will be useful in Section \ref{sec:3scalarshol} when we discuss the inflationary consistency relation for the scalar bispectrum \cite{Maldacena:2002vr}.

\subsection{Introducing the stress tensor}
\label{sec:Tintro}

In this subsection, we now generalise our discussion of conformal perturbation theory to include the stress tensor.
After first establishing the definition of this operator in the perturbed theory, we return to the unperturbed CFT to 
consider the form of 3-point correlators with mixed scalar and stress tensor insertions.  From these correlators we may read off the corresponding OPEs, and hence understand the behaviour of stress tensor insertions in correlators of the perturbed theory.

Working henceforth in renormalised perturbation theory, the action takes the form
\[
S = S_{CFT}+\int\d^3 x \sqrt{g}\phi \O,
\]
and we will assume the two sources $\phi$ and $g_{ij}$ are functionally independent of one another.
Now, in a completely general theory, the renormalised scalar operator $\O$ may depend on either of the sources $\phi$ and $g_{ij}$.
Such a dependence, however, generally introduces additional nearly marginal scalar operators into the spectrum,\footnote{Unless it is possible to re-express these operators in terms of $\O$ and $T_{ij}$ at higher order in $\lambda$.}
 e.g., $\delta\O/\delta\phi$ with dimension $3-2\lambda$ or $g^{ij}\delta\O/\delta g^{ij}$ with dimension $3-\lambda$.
From a bulk perspective, this would then correspond to the introduction of additional light scalar fields besides the inflaton resulting in a multi-scalar model.
Since our present aim is to concentrate on the single-field case, we will assume that $\O$ is independent of the sources $\phi$ and $g_{ij}$, at least to the leading order in $\lambda$ at which we work.  

In this case, the stress tensor $T_{ij}$ in the perturbed theory is related to the stress tensor $\T_{ij}$ in the unperturbed CFT according to
\[\label{Tdef}
T_{ij} = \T_{ij} -\phi\O g_{ij}.
\]
It follows that the transverse traceless piece of these stress tensors is then identical.  
Defining the transverse traceless projector 
\[\label{proj_op_x}
\Pi_{ijkl}=\frac{1}{2}\big(\pi_{ik}\pi_{jl}+\pi_{il}\pi_{jk}-\pi_{ij}\pi_{kl}\big), 
\]
where $\pi_{ij}=\delta_{ij}-\p^{-2}\p_i\p_j$ is the transverse projector on a flat background, we have
$
T^\perp_{ij} = \T^\perp_{ij}
$
where $T^\perp_{ij}\equiv\Pi_{ijkl}T_{kl}$ and $\T^{\perp}_{ij}$ is defined similarly. 
In momentum space, where $\pi_{ij}=\delta_{ij}-q_iq_j/q^2$, we may equivalently write $T^{(s)}(q)=\T^{(s)}(q)$ where the helicity projection
\[
T^{(s)}(q) = \frac{1}{2}\ep_{ij}^{(s)}(-q)T_{ij}(q),
\]
and similarly for $\T^{(s)}(q)$. Here, the factor of one half arises because our helicity tensors $\ep_{ij}^{(s)}(q)$ satisfy the identities
\[
\label{PiTTdecomp}
\Pi_{ijkl}(q) = \half\ep^{(s)}_{ij}(q)\ep^{(s)}_{kl}(-q), \qquad
 \ep^{(s)}_{ij}(q)\ep^{(s')}_{ij}(-q) = 2\delta^{ss'},
\]
where the helicities take values $s=\pm 1$ and repeated indices are to be summed.

In an exact CFT, the 2-point stress tensor correlator takes the form \cite{Osborn:1993cr}
\[\label{CFT_TT_x}
\<\T_{ij}(x_1)\T_{kl}(x_2)\>_0 =  \frac{\alpha_T}{|x_{12}|^6}\, \I_{ij,kl}(x_{12}),
\]
where
\[
\I_{ij,kl}(x) = \frac{1}{2}\big(I_{ik}(x)I_{jl}(x)+I_{il}(x)I_{jk}(x)\big)-\frac{1}{3}\delta_{ij}\delta_{kl}, \qquad
I_{ij}(x) = \delta_{ij}-\frac{2x_ix_j}{x^2},
\]
while the 3-point correlators we will need are
\begin{align}\label{CFT_OOT}
\<\O(x_1)\O(x_2)\T_{ij}(x_3)\>_0 &= \frac{\tilde{C}}{|x_{12}|^{3-2\lambda}|x_{23}|^3|x_{31}|^3}\,t_{ij}(X), \\
\label{CFT_TTO}
\<\T_{ij}(x_1)\T_{kl}(x_2)\O(x_3)\>_0 &= \frac{\alpha_{TT}}{|x_{12}|^{3+2\lambda}|x_{23}|^{3-2\lambda}|x_{31}|^{3-2\lambda}}\,
\I_{ij,mn}(x_{31})\I_{kl,rs}(x_{23})t_{mn,rs}(X).
\end{align}
In these formulae,\footnote{The specific coefficients appearing here derive from solving (3.6) and (6.20) in \cite{Osborn:1993cr} at leading order in $\lambda$, where $c$ in (3.4) of \cite{Osborn:1993cr} equals $\alpha_{TT}$ here.}
\begin{align} \label{tX_def}
X_i &= -\frac{x_{31i}}{x_{31}^2}-\frac{x_{23i}}{x_{23}^2}, \qquad
t_{ij}(X) = \frac{X_i X_j}{X^2}-\frac{1}{3}\delta_{ij}, \qquad \tilde{C} = -\frac{9\alpha}{8\pi}+O(\lambda), \\
t_{ij,kl}(X) &= -5 t_{ij}(X)t_{kl}(X)-t_{ik}(X)\delta_{jl}-t_{jl}(X)\delta_{ik}+\frac{4}{3}t_{ij}(X)\delta_{kl}+\frac{4}{3}t_{kl}(X)\delta_{ij}\nn\\
&\quad +\frac{1}{3}\delta_{ik}\delta_{jl}+\delta_{il}\delta_{jk}-\frac{2}{9}\delta_{ij}\delta_{kl} + O(\lambda),
\end{align}
and we note in particular that the overall normalisation $\tilde{C}$ of the 3-point correlator with a single stress tensor insertion is fixed by the trace Ward identity \cite{Cardy:1987dg}.  

Expanding out the 3-point correlators in the limit when two insertion points coincide, we obtain the following OPE contributions
with scaling dimensions close to three,
\begin{align}
\label{OT_OPE}
\T_{ij}(x_1)\O(x_2) &= A_{ij}(x_{12})\O(x_2)+ B_{ijkl}(x_{12})\T_{kl}(x_2) + \ldots, \\ 
\label{OO_OPE2}
\O(x_1)\O(x_2) &= \frac{C}{|x_{12}|^{3-\lambda}}\,\O(x_2)+ C_{ij}(x_{12})\T_{ij}(x_2) + \ldots,
\end{align}
where\footnote{To rewrite $A_{ij}$ as a well-defined distribution over $\mathbb{R}^3$ one may use differential regularisation as discussed in \cite{Osborn:1993cr}. The remaining OPE coefficients $B_{ij}$ and $C_{ij}$ already have well-defined Fourier transforms.
}
\begin{align}
\label{Aij_def}
A_{ij}(x) &= -\frac{9\alpha}{8\pi x^3}\Big(\frac{x_i x_j}{x^2}-\frac{1}{3}\delta_{ij}\Big)+O(\lambda), \\[1ex]
\label{Bij_def}
B_{ijkl}(x) &= \frac{\alpha_{TT}}{\alpha_T x^{3-\lambda}}\Big({-}\frac{5}{x^4}x_ix_jx_kx_l{+}\frac{7}{3x^2}x_kx_l\delta_{ij}
{-}\frac{1}{x^2}x_k x_{(i}\delta_{j)l}{-}\frac{1}{x^2}x_l x_{(i}\delta_{j)k}{+}\delta_{k(i}\delta_{j)l}\Big)+O(\lambda),\\[1ex]
\label{Cij_def}
C_{ij}(x) &= \frac{9\alpha}{8\pi \alpha_T}\frac{1}{x^{3-2\lambda}}\Big(\frac{x_ix_j}{x^2}-\frac{1}{3}\delta_{ij}\Big)+O(\lambda).
\end{align}

Our first task is now to verify that the presence of the stress tensor in the $\O\O$ OPE does not modify our earlier computations of  the scalar 2- and 3-point functions.  Fortunately, this is indeed the case.  To illustrate this, let us consider the regulated integral $\I_1$ given in \eqref{I1def}.  
Varying with respect to the cutoff $\Lambda$ and using the OPE \eqref{OO_OPE2}, the r.h.s.~of our earlier result \eqref{I1_ODE}
acquires a new contribution
\begin{align}
&-\int\d^3 z_1 \big[C_{ij}(z-x_1)\<\T_{ij}(x_1)\O(x_2)\>_0 \delta(|z-x_1|-\Lambda) 
+C_{ij}(z-x_2)\<\O(x_1)\T_{ij}(x_2)\>_0\delta(|z-x_2|-\Lambda)\big] \nn\\&
= -[\<\T_{ij}(x_1)\O(x_2)\>_0+\<\O(x_1)\T_{ij}(x_2)\>_0]\int \d^3 y\, C_{ij}(y) \delta(|y|-\Lambda).
\end{align}
Now, in this particular example, the correlators out the front happen to vanish, but in general this will not be the case when we start from the integral $\I_n$ containing $n$ integrated scalar insertions.  Rather, the point is that the residual correlators factor out leaving the integral over a spherical shell of the OPE coefficient $C_{ij}$.  From \eqref{Cij_def}, this OPE coefficient is isotropic and traceless, and so its integral over a spherical shell simply vanishes.
We thus obtain no new corrections to our earlier results for the scalar 2- and 3-point functions.\footnote{In fact, the isotropy of $C_{ij}$ alone would be sufficient to establish this, since we would obtain residual correlators involving the trace of the stress tensor which vanish by the trace Ward identities, noting that in the regulated correlators none of the insertion points are coincident.}  

Let us next consider how to evaluate correlators in the perturbed theory involving one or more fixed insertions of the stress tensor.  Repeating our above argument, we cannot contract an integrated scalar insertion with a fixed stress tensor to generate a fixed scalar insertion, since the OPE coefficient $A_{ij}$ in \eqref{Aij_def} is likewise isotropic and traceless.  Even if $A_{ij}$ were not isotropic and traceless, its scaling as $|x|^{-3}$ means that any resulting correlator would be suppressed by a factor of $\lambda$.  Specifically, recall that each integrated scalar insertion obtained by expanding $\<\exp(-\int\vphi\Lambda^{-\lambda}\O)\>_0$ carries a factor of $\vphi$ which contributes one power of $\lambda$.  In the case we considered in Section \ref{sec:scalar2pt} (namely, contracting a fixed scalar insertion with an integrated scalar insertion to generate a fixed scalar insertion), this factor of $\lambda$ was offset by a factor of $1/\lambda$ arising from integrating the OPE coefficient $C/|x|^{3-\lambda}$ over a spherical shell, as we saw in \eqref{I1_ODE} and \eqref{I1eq}.  In the present case, however, the scaling of $A_{ij}$ as $|x|^{-3}$ means that we do not acquire this compensating factor, hence any fixed scalar insertion obtained from the contraction of an integrated scalar insertion with a fixed stress tensor 
will be suppressed by a factor of $\lambda$ relative to leading order.
For this same reason we may also ignore operators in the OPE with scaling dimensions not close to three.

In principle, the contraction of an integrated scalar insertion with a fixed stress tensor insertion may generate a fixed stress tensor insertion, since the OPE coefficient $B_{ijkl}$ in \eqref{Bij_def} scales as $1/|x|^{3-\lambda}$ and the integral over the spherical shell will not in general vanish.  
In fact, however, $\alpha_{TT}$ (and hence $B_{ijkl}$) vanishes for the CFT dual to Einstein gravity.  
To see this, note first that $\alpha_{TT}$ is a property of the UV CFT alone (as opposed to the full perturbed theory), and so may be extracted from an AdS/CFT calculation on an exact AdS background.  For the CFT correlator $\<\T_{ij}\T_{kl}\O\>_0$ to be nonzero we would require a nonvanishing graviton-graviton-scalar coupling in the expansion of the bulk action about this background.
Given a bulk action of the form
\[
\frac{1}{2\kappa^2}\int d^4x \sqrt{-g}[R-(\p\Phi)^2-2\kappa^2 V(\Phi)],
\]
perturbing about an AdS background involves setting $\Phi=\vphi_0+\delta\vphi$, where $\vphi_0$ is a constant.  A graviton-graviton-scalar vertex may then only come from the expansion of $\sqrt{-g}V(\Phi)$, yet since the background is a solution of the scalar field equation of motion this term is a tadpole and vanishes.\footnote{This may also be seen from the explicit calculation of the cubic interaction terms in \cite{Maldacena:2002vr}.  When the time derivative of the background scalar field vanishes we must use the gauge (3.2); the graviton-graviton-scalar vertex is then given by the first line of (3.17), which vanishes after taking into account (3.11).}    
For the CFT dual to Einstein gravity then, the correlator $\<\T_{ij}\T_{kl}\O\>_0$ and hence $\alpha_{TT}$ must vanish.
In consequence one cannot generate a fixed stress tensor insertion from contracting an integrated scalar insertion with a fixed stress tensor insertion, at least at leading order in $\lambda$.  (At higher order, however, this should still be possible in order to generate a nontrivial momentum dependence in the tensor 2-point function, and hence a nonvanishing tilt in the inflationary tensor power spectrum through the holographic formula \eqref{DeltaST}.)  
We stress also that this conclusion is specific to Einstein gravity; for more general bulk actions it may be possible to obtain a nonvanishing $\alpha_{TT}$ which would allow integrated scalars to contract with fixed stress tensors. 
It would be interesting to explore this further in specific models.

In summary then, for Einstein gravity at leading order in $\lambda$ we find that insertions of the stress tensor are essentially inert.  Contractions of integrated scalars with fixed stress tensors produce no contribution, and contractions of integrated scalars cannot generate stress tensor insertions.  Thus, only scalar insertions participate in the leading order resummation process.

\subsection{Stress tensor correlators}
\label{sec:stresscorrs}

Having ascertained the rules of the resummation process when stress tensor insertions are included,
it remains to evaluate the specific 2- and 3-point correlators that will appear in our holographic formulae.  No new methods are required for this analysis, only a straightforward application of those developed above for correlators with scalar insertions.

Beginning with the 2-point functions, at leading order in $\lambda$ we find that
\begin{align} 
\label{2pts_in_x}
\<T^\perp_{ij}(x_1)T^\perp_{kl}(x_2)\>&=\<\T^\perp_{ij}(x_1)\T^\perp_{kl}(x_2)\>_0, \\
\label{OTperpdef}
\<T^\perp_{ij}(x_1)\O(x_2)\> &= \Big[1+\frac{\phi}{\vphi_1}|x_{12}|^\lambda\Big]^{-2}\<\T^\perp_{ij}(x_1)\O(x_2)\>_0 = 0.
\end{align}
The first of these results is a straightforward reflection of the fact that integrated scalar insertions yield no contribution when contracted with fixed stress tensor insertions.   As for the second result, although we may contract each of the integrated scalar insertions against the single fixed scalar insertion, each contraction simply returns a fixed scalar insertion and so, after resumming the binomial series to obtain the middle expression, we arrive at a pure CFT 2-point function of two operators with mismatched dimensions which vanishes.  

Let us now consider the conversion of these results to momentum space.  The trace and diffeomorphism Ward identities imply that the stress tensor 2-point function of a CFT is both transverse and traceless, and hence must take the form
\[
\lla \T_{ij}(q)\T_{kl}(-q)\rra_0 = A_0(q) \Pi_{ijkl},
\]
where the transverse traceless projector $\Pi_{ijkl}$ is defined in \eqref{proj_op_x}.
Projected into a helicity basis, this expression is equivalent to
\[
\lla \T^{(s)}(q)\T^{(s')}(-q)\rra_0 = \frac{1}{2} A_0(q)\delta^{ss'}.
\]
To identify the coefficient $A_0(q)$ explicitly, we Fourier transform the position space expression \eqref{CFT_TT_x} with contracted indices to give
\[
\lla \T_{ij}(q)\T_{ij}(-q)\rra_0 = \frac{\pi^2}{12}\alpha_T q^3 = 2 A_0(q).
\]
Thus, in momentum space the results \eqref{2pts_in_x} read
\begin{align}
\label{TsTs}
\lla T^{(s)}(q)T^{(s')}(-q)\rra &= \lla \T^{(s)}(q)\T^{(s')}(-q)\rra_0 = \frac{\pi^2}{48}\alpha_T q^3 \delta^{ss'}, \\
\label{TsO}
\lla T^{(s)}(q)\O(-q)\rra &= 0.
\end{align}

Focusing next on the 3-point functions, in the case where all the $|x_{ij}|^\lambda$ are comparable, at leading order in $\lambda$  we
find
\begin{align}
\<\O(x_1)\O(x_2)T^\perp_{ij}(x_3)\> &= \Big[1+\frac{\phi}{\vphi_1}|L|^\lambda\Big]^{-4} \<\O(x_1)\O(x_2)\T^\perp_{ij}(x_3)\>_0. 
\end{align}
The power featuring in the prefactor is the same as that for the scalar 2-point function since in both cases we are resumming the binomial series resulting from contracting against two fixed scalar insertions.
Through consideration of the various limiting cases using the OPEs \eqref{OT_OPE} and \eqref{OO_OPE2} as well as the 2-point results  \eqref{2pt_result}, \eqref{2pts_in_x} and \eqref{OTperpdef}, we may further refine this to
\begin{align}\label{OOTformula1}
\<\O(x_1)\O(x_2)T^\perp_{ij}(x_3)\> &= \Big[1+\frac{\phi}{\vphi_1}|x_{12}|^\lambda\Big]^{-4} \<\O(x_1)\O(x_2)\T^\perp_{ij}(x_3)\>_0. \end{align}
To convert this result to momentum space, it is useful to first re-express it as a sum of exact CFT 3-point functions with shifted dimensions.
From the explicit form of the CFT correlator in \eqref{CFT_OOT}, it follows that to leading order 
\[
|x_{12}|^{2n\lambda} \<\O_\Delta(x_1)\O_\Delta(x_2)\T_{ij}(x_3)\>_0 = \<\O_{\Delta-n\lambda}(x_1)\O_{\Delta-n\lambda}(x_2)\T_{ij}(x_3)\>_0.
\]
The result \eqref{OOTformula1} may then be cast in the desired form, 
\[\label{sum_formula_OOT}
\<\O_\Delta(x_1)\O_\Delta(x_2)\T^\perp_{ij}(x_3)\> = \frac{1}{6}\sum_{n=0}^\infty (n+3)(n+2)(n+1)\Big({-}\frac{\phi}{\vphi_1}\Big)^n\<\O_{\Delta_n}(x_1)\O_{\Delta_n}(x_2)\T^\perp_{ij}(x_3)\>_0,
\]
where $\Delta_n=\Delta-n\lambda/2$.  After evaluating the Fourier transform of the 
exact CFT correlator \eqref{CFT_OOT}, the perturbed correlator \eqref{OOTformula1} in momentum space may be obtained by summing this series.

The general result, derived in Appendix \ref{app:OOT_FT}, may take one of two forms according to our choice of nonzero reference momentum (namely, whether we select the momentum $q_3$ associated with $T_{ij}$, or else one of the momenta associated with $\O$, say $q_1$). Each form covers one (but not both) of the two distinct squeezed limits $q_1\tto 0$ and $q_3\tto 0$, as well as the common case where all the $q_i^{-\lambda}$ are comparable, with both forms agreeing in this latter case.
For comparison with the standard inflationary results, however, it is sufficient to examine each of these various cases separately.

Firstly, for the quasi-equilateral case in which all three momenta are comparable, namely
\[\label{quasi_eq}
q_1^{-\lambda}(1+O(\lambda))=q_2^{-\lambda}(1+O(\lambda))=q_3^{-\lambda}(1+O(\lambda)),
\]
we find that 
\[
\<\!\<\O_{\Delta}(q_1)\O_{\Delta}(q_2)\T^\perp_{ij}(q_3)\>\!\> = \A_{eq}(q_1,q_2,q_3) \Pi_{ijkl}(q_3)q_{1k}q_{1l},
\]
where
\[\label{Aeq}
\A_{eq}(q_1,q_2,q_3) = \frac{\alpha\pi^2}{6} q_3^{-2\lambda} \Big[1+\frac{\phi}{\vphi_1} q_3^{-\lambda}\Big]^{-4} \frac{(-a_{123}^3+a_{123}b_{123}+c_{123})}{a_{123}^2}
\]
with the elementary symmetric polynomials 
\begin{align}
\label{sym_polys}
&a_{123} = q_1+q_2+q_3, \qquad b_{123} = q_1q_2+q_2q_3+q_3q_1, \qquad c_{123}= q_1q_2q_3.
\end{align}
In a helicity basis this may be re-expressed as\footnote{See Appendix A of \cite{Bzowski:2011ab} for explicit formulae converting between tensor and helicity bases.} 
\[\label{OOTs_eq_FT}
\<\!\<\O_{\Delta}(q_1)\O_{\Delta}(q_2)\T^{(s_3)}(q_3)\>\!\> = \frac{J^2}{8\sqrt{2}\,q_3^2} \A_{eq}(q_1,q_2,q_3),
\]
where
\begin{align}
\label{J_def}
 \J^2 &=(q_1+q_2+q_3)(-q_1+q_2+q_3)(q_1-q_2+q_3)(q_1+q_2-q_3) \nn\\
&= -a_{123}(a_{123}^3-4a_{123}b_{123}+8c_{123}).
\end{align}
(From a geometrical perspective, $\J$ then represents a quarter of the area of the triangle with
side lengths $q_1$, $q_2$ and $q_3$ according to Heron's formula.)

The remaining cases are then the two squeezed limits $q_1\tto 0$ and $q_3\tto 0$, where from our results in Appendix \ref{app:OOT_FT} we find that the correlator in the perturbed theory vanishes at leading order
\[\label{OOT_sq1}
\<\!\<\O(0)\O(q)T^{(s)}(-q)\>\!\> = 0, \qquad \<\!\<\O(q)\O(-q)T^{(s)}(0)\>\!\> = 0.
\]
As a consistency check on the first of these, we may confirm that 
\[
\<\!\<\O(0)\O(q)T^{(s)}(-q)\>\!\>= -\frac{\p}{\p\phi}\<\!\<\O(q)T^{(s)}(-q)\>\!\> = 0.
\]

The final correlator in the perturbed theory that we need to evaluate is the 3-point function with two stress tensor insertions and one scalar.  Following arguments analogous to those we have used above, we obtain
\[\label{TTOcorr1}
\<\!\<T^\perp_{ij}(x_1)T^\perp_{kl}(x_2)\O(x_3)\>\!\> = \Big[1+\frac{\phi}{\vphi_1}|x_{12}|^\lambda\Big]^{-2}
\<\!\<\T^\perp_{ij}(x_1)\T^\perp_{kl}(x_2)\O(x_3)\>\!\>_0.
\]
As we saw in the previous subsection, however, for the CFT dual to Einstein gravity $\alpha_{TT}$ is zero hence the correlator on the r.h.s.~vanishes.  In the perturbed theory, therefore, the 3-point correlator of two stress tensors and one scalar vanishes at leading order in $\lambda$.  We thus have
\[
\<\!\<T^{(s_1)}(q_1)T^{(s_2)}(q_2)\O(q_3)\>\!\> = 0.
\]

To summarise the main results in this section, for 2-point functions we found \eqref{TsTs} and \eqref{TsO}, while for the 3-point function of two scalars and one stress tensor in the quasi-equilateral momentum configuration, we found \eqref{OOTs_eq_FT} with $\A_{eq}$ given in \eqref{Aeq}. The 3-point correlator of two stress tensors and one scalar, however, vanishes at leading order.

\section{Holographic calculation of inflationary correlators} 
\label{sec:hol_calc}

Having now computed all the correlators of interest to us, let us briefly recount the main features of the holographic framework for cosmology developed in \cite{McFadden:2009fg, McFadden:2010na, McFadden:2010vh, McFadden:2011kk}.
This framework is based on standard AdS holography and the domain-wall/cosmology correspondence \cite{Cvetic:1994ya,Skenderis:2006jq, McFadden:2010na, McFadden:2010vh, McFadden:2011kk}, a one-to-one mapping between perturbed cosmological solutions and the perturbed asymptotically AdS domain-wall spacetimes which describe holographic RG flows.  
The domain-wall/cosmology correspondence may be expressed as a simple analytic continuation taking a solution of the perturbed domain-wall equations of motion to a solution of the perturbed cosmological equations of motion.  In fact, all that is required is to invert the inflationary potential $\kappa^2 V\tto -\kappa^2 V$ while continuing $q \tto -iq$, where $q=\sqrt{\vec{q}\,^2}$ is the magnitude of the 3-momentum on spatial slices for the perturbations.  The sign in this last continuation is chosen so as to map perturbed domain-wall solutions that are regular in the interior to cosmological perturbations originating in the Bunch-Davies vacuum. 
An explicit proof of this mapping at quadratic order in gauge-invariant perturbation theory may be found in \cite{McFadden:2011kk}.

We may equally well express this bulk analytic continuation in the language of the dual QFT.  The continuation $q\tto -iq$ for the magnitude of the momentum is the same, while for an $SU(N)$ gauge theory with fields in the adjoint representation, inverting the inflationary potential amounts to sending $N^2\tto -N^2$.  (For the $O(N)$ vector model in the proposed Vasiliev dS/CFT correspondence \cite{Anninos:2011ui}, the analogous continuation is instead $N\tto -N$.)
From an observational perspective, 
we are primarily interested in perturbatively quantising small fluctuations about an FRW background, which in the dual QFT corresponds to working within large-$N$ perturbation theory.
To compute tree-level cosmological correlators, it is therefore sufficient to evaluate QFT correlators at leading order in $1/N$ in the {\it ordinary} QFT dual to the domain-wall spacetime, and then apply the analytic continuation directly to these correlators.  Inflationary loop corrections are similarly expected to map to subleading $1/N$ corrections in the dual QFT.  Understanding holographic cosmology beyond large-$N$ perturbation theory remains an open problem.

\subsection{Holographic formulae}

Exact holographic formulae expressing tree-level cosmological 2- and 3-point functions in terms of correlation functions of the dual QFT were derived in \cite{McFadden:2009fg, McFadden:2010vh, McFadden:2011kk}.  Specifically, these holographic formulae relate cosmological correlators to correlators of the stress tensor in the dual QFT.  In the present case, these formulae may be simplified further using the Ward identities derived in Appendix \ref{app:Ward} to replace insertions of the trace of the stress tensor with insertions of $\O$.  Moreover, when working to leading order in $\lambda$, the analytic continuations appearing in the formulae of \cite{McFadden:2009fg, McFadden:2010vh, McFadden:2011kk} simply amount to a few changes of sign and may be trivially implemented.\footnote{Explicitly, for adjoint fields all correlators are proportional to $N^2$, where $N$ is the rank of the gauge group.  The continuation $N^2\tto -N^2$ then contributes an overall sign.  The continuation $q\tto -iq$ on the magnitude of the momenta sends, e.g., $q^{3-2\lambda}\tto i q^{3-2\lambda}(1+O(\lambda))$, since $(-i)^\lambda = 1+O(\lambda)$.  As the underlying CFT correlators are homogeneous functions of $q$ whose overall scaling is $q^{3+p\lambda}$ for some integer $p$, this second continuation merely contributes an overall factor of $i$ to correlators, which is eliminated upon taking the imaginary part as directed by the holographic formulae.}  We summarise the simplified holographic formulae obtained by these elementary manipulations below.

The leading order cosmological 2-point functions are given by
\begin{align}
\<\!\<\z(q)\z(-q)\>\!\> = \frac{1}{2\lambda^2\phi^2 \lla\O(q)\O(-q)\rra}, \qquad
\lla \g^{(s)}(q)\g^{(s')}(-q)\rra = \frac{\delta^{ss'}}{ A(q) },
\end{align}
where $\z$ is the comoving curvature perturbation and $\g^{(s)}$ the graviton (see \cite{McFadden:2011kk} for precise definitions) and
\[\label{tensorAdef}
\lla T^{(s)}(q)T^{(s')}(-q)\rra = \frac{1}{2} A(q)\delta^{ss'},
\]
from which we obtain the power spectra
\[\label{DeltaST}
\Delta_S^2(q) = \frac{q^3}{4\pi^2 \lambda^2\phi^2 \lla\O(q)\O(-q)\rra}, \qquad
\Delta_T^2(q) = \frac{2q^3}{\pi^2 A(q) }.
\]

The cosmological 3-point functions at leading order in $\lambda$ are 
\begin{align}
\label{zzz}
&\lla \z(q_1) \z(q_2) \z(q_3) \rra  
 =   \frac{ \phi \lla \mathcal{O}(q_1) \mathcal{O}(q_2) \mathcal{O}(q_3) \rra - \sum_{j=1}^3 \lla \mathcal{O}(q_j) \mathcal{O}(-q_j) \rra }{4\lambda^3 \phi^4 \prod_{j=1}^3  \lla \mathcal{O}(q_j) \mathcal{O}(-q_j) \rra}, \\[2ex]
 \label{zzg}
&\lla \zeta(q_1) \zeta(q_2) \gamma^{(s_3)}(q_3) \rra 
 =  \frac{-\lambda \phi \lla \mathcal{O}(q_1) \mathcal{O}(q_2) T^{(s_3)}(q_3) \rra 
-\lla \mathcal{O}(q_3) T^{(s)}(-q_3) \rra 
}{2\lambda^3 \phi^3  \lla \mathcal{O}(q_1) \mathcal{O}(-q_1) \rra  \lla \mathcal{O}(q_2) \mathcal{O}(-q_2) \rra  A(q_3)}, \\[2ex]  
\label{zgg}
&\lla \zeta(q_1) \gamma^{(s_2)}(q_2) \gamma^{(s_3)}(q_3) \rra \nn\\[1ex]&
=  \frac{
   \lla \mathcal{O}(q_1) T^{(s_2)}(q_2) T^{(s_3)}(q_3) \rra 
+ \frac{1}{4} \lambda \phi \Theta^{(s_2 s_3)} \lla \mathcal{O}(q_1) \mathcal{O}(q_1) \rra 
-  2  \lla \mathcal{O}(q_1) Y_T^{(s_2 s_3)}(-q_1) \rra 
}{\lambda \phi  \lla \mathcal{O}(q_1) \mathcal{O}(q_1) \rra  A(q_2)  A(q_3)} , 
\end{align}
where in this last formula, 
\[
\Theta^{(s_2s_3)} = \pi_{ij}(q_1)\ep_{ik}^{(s_2)}(-q_2)\ep_{kj}^{(s_3)}(-q_3).
\] 
An explicit expression for this quantity in terms of the momenta may be found in \cite{Bzowski:2011ab}, although we will not need it here. We will return to the definition of $Y_T^{(s_2s_3)}$ shortly in Section \ref{non_g_hol}.

The 2-point terms in the numerators of the holographic formulae above are important, despite the fact they represent semi-local contact terms, i.e., terms that are non-analytic in only one of the three momentum magnitudes, or equivalently in position space, terms that contribute only when two of the three insertion points are brought together.  Due to the product of three 2-point functions in the denominator of the holographic formulae, the 2-point terms in the numerator produce a net contribution to the cosmological correlator that is non-analytic in {\it two} momenta, and hence may contribute to so-called local-type non-Gaussianity.  It is therefore essential to correctly compute and retain these semi-local contributions in the numerator of the holographic formulae; we refer the reader to \cite{McFadden:2010vh, McFadden:2011kk, Bzowski:2011ab} for their derivation along with further discussion.

\subsection{Power spectra}

Inserting \eqref{OO_FT} into the holographic formula \eqref{DeltaST}, we find the scalar power spectrum
\[\label{hol_power_spec}
\Delta_S^2(q) = \frac{3}{\pi^4\alpha\lambda^2\phi^2}\,q^{2\lambda}\Big[1+\frac{\phi}{\vphi_1}q^{-\lambda}\Big]^4.
\]

Referring back to \eqref{renphi}, we see the renormalised coupling $\phi$ is essentially arbitrary as it depends on the initial condition $\vphi(\Lambda_0)=\vphi_0$ when we integrated the $\beta$-function.  This arbitrariness corresponds to our freedom to rescale the operator $\O$ in the perturbed action. It is therefore useful to repackage this arbitrariness into a momentum scale $q_0$ defined by
\[\label{q0def}
q_0^{-\lambda} \equiv \frac{\vphi_1}{\phi}=\Big(\frac{\vphi_1}{\vphi_0}-1\Big)\Lambda_0^\lambda,
\]
in terms of which the power spectrum \eqref{hol_power_spec} may be rewritten as
\[\label{hol_power_spec2}
\Delta_S^2(q) = \frac{1}{16}\Delta_S^2(q_0) \Big(\frac{q}{q_0}\Big)^{-2\lambda}\Big[1+\Big(\frac{q}{q_0}\Big)^\lambda\Big]^4,
\qquad \Delta_S^2(q_0) = \frac{192\, C^2}{\pi^2\alpha \lambda^4}.
\]
The 2-point function for curvature perturbations is accordingly
\[
\<\!\<\z(q)\z(-q)\>\!\> = \frac{24\, C^2}{\alpha \lambda^4 q^3}\Big(\frac{q}{q_0}\Big)^{-2\lambda}\Big[1+\Big(\frac{q}{q_0}\Big)^\lambda\Big]^4.
\]

The spectral tilt 
\[\label{spec_tilt}
n_S-1 \equiv \frac{\d \ln \Delta_S^2(q)}{\d \ln q} = 2\lambda-4\lambda\Big[1+\Big(\frac{q}{q_0}\Big)^{\lambda}\Big]^{-1},
\]
and so for small momenta $q/q_0\ll 1$ we obtain a red-tilted spectrum with $n_S-1 \approx -2\lambda$, while for large momenta $q/q_0\gg 1$ we obtain a blue tilt  $n_S-1 \approx 2\lambda$.  This behaviour reflects the fact that $\O$ is marginally relevant in the UV  but marginally irrelevant in the IR, i.e., $\Delta_{UV}=3-\lambda$ while $\Delta_{IR}=3+\lambda$.

The tensor power spectrum at leading order in $\lambda$ may be found by inserting \eqref{TsTs} into \eqref{DeltaST}, giving
\[\label{DeltaT_result}
\Delta_T^2(q) = \frac{48}{\pi^4 \alpha_T}
\]
or equivalently
\[
\lla \g^{(s)}(q)\g^{(s')}(-q)\rra = \frac{24}{\pi^2\alpha_T q^3}\,\delta^{ss'}.
\]
The exact scale invariance of the tensor power spectrum at leading order in $\lambda$ is due to the vanishing of $\alpha_{TT}$ and thus the correlator $\<\T^{(s)}\T^{(s')}\O\>_0$ for the CFT dual to Einstein gravity.

\subsection{Non-Gaussianities}
\label{non_g_hol}

\subsubsection{Three scalars}
\label{sec:3scalarshol}

Inserting our earlier results \eqref{OO_FT} and \eqref{OOO_FT} into \eqref{zzz} 
and using \eqref{q0def}, we obtain
\begin{align}
\<\!\<\z(q_1)\z(q_2)\z(q_3)\>\!\> &= 
\frac{576\, C^4}{\alpha^2 \lambda^7}
\Big({-}1+2\Big[1+\Big(\frac{q_3}{q_0}\Big)^{\lambda}\Big]^{-1}\Big)
\nn\\&\qquad\times
\sum_{i<j} q_i^{-3}q_j^{-3} \Big(\frac{q_i}{q_0}\Big)^{-2\lambda}\Big(\frac{q_j}{q_0}\Big)^{-2\lambda}
\Big[1+\Big(\frac{q_i}{q_0}\Big)^{\lambda}\Big]^4\Big[1+\Big(\frac{q_j}{q_0}\Big)^{\lambda}\Big]^4, 
\end{align}
or equivalently,
\[\label{zzz_hol}
\<\!\<\z(q_1)\z(q_2)\z(q_3)\>\!\>  
= -\frac{1}{2}(n_S(q_3)-1)\sum_{i<j}
\<\!\<\z(q_i)\z(-q_i)\>\!\>\<\!\<\z(q_j)\z(-q_j)\>\!\>,
\]
where $q_3$ is the nonzero reference momentum chosen as the largest of the three momenta.
Thus, at leading order in $\lambda$, we obtain a scalar bispectrum of purely local form.

To set up a comparison with the standard inflationary results in Section \ref{sec:nGcosmozzz} it is useful to record here that
in the quasi-equilateral case \eqref{quasi_eq} where all the $q_i^{-\lambda}$ are comparable, \eqref{zzz_hol} reduces to
\[\label{zzz_hol_eq}
\<\!\<\z(q_1)\z(q_2)\z(q_3)\>\!\> = 
\frac{576\, C^4}{\alpha^2 \lambda^7}
\Big({-}1+2\Big[1+\Big(\frac{q_3}{q_0}\Big)^{\lambda}\Big]^{-1}\Big)
\Big(\frac{q_3}{q_0}\Big)^{-4\lambda}
\Big[1+\Big(\frac{q_3}{q_0}\Big)^{\lambda}\Big]^8 \sum_{i<j}q_i^{-3}q_j^{-3},
\]
where the interpretation of the terms in the prefactor will become transparent in Section \ref{sec:nGcosmozzz}.

In the squeezed limit where, e.g., $q_1$ becomes small, the leading behaviour of \eqref{zzz_hol} is
\[\label{sqzzzinx}
\lim_{q_1\tto 0} \<\!\<\z(q_1)\z(q_2)\z(q_3)\>\!\> \approx -(n_S(q_3)-1)
\<\!\<\z(q_3)\z(-q_3)\>\!\>\<\!\<\z(q_1)\z(-q_1)\>\!\>
\]
in accordance with the consistency relation of \cite{Maldacena:2002vr}.
Indeed, a simple yet independent way to understand this behaviour is to observe that 
\begin{align}
-(n_S(q)-1)\lla \O(q)\O(-q)\rra  &= -\lla \O(q)\O(-q)\rra\frac{\p}{\p \ln q}\ln \Delta_S^2(q) = \Big({-}3+\frac{\p}{\p \ln q}\Big)\lla \O(q)\O(-q)\rra \nn\\&
= \lambda\phi\lla \O(0)\O(q)\O(-q)\rra-2\lambda \lla \O(q)\O(-q)\rra ,
\end{align}
where in the first line we used the holographic formula for the power spectrum \eqref{DeltaST} and in the second line we used our earlier relation \eqref{sqformOOO}.
Thus, when $q_1 \ll q_2\approx q_3$, the leading behaviour of ($\lambda$ times) the numerator of the holographic formula \eqref{zzz} is
\[
 \lambda\phi\lla \O(q_1)\O(q_2)\O(q_3)\rra  -\sum_{i=1}^3\lambda \lla \O(q_i)\O(-q_i)\rra
 \approx -(n_S(q_3)-1)\lla\O(q_3)\O(-q_3)\rra ,
\]
since $\lla \O(q_1)\O(-q_1)\rra \ll \lla\O(q_3)\O(-q_3)\rra$.   Meanwhile, the leading behaviour of ($\lambda$ times) the denominator of the holographic formula \eqref{zzz} is
\[
\frac{1}{4\lambda^4\phi^4 \lla \O(q_1)\O(-q_1)\rra \lla \O(q_3)\O(-q_3)\rra ^2 } = \frac{1}{\lla \O(q_3)\O(-q_3)\rra}\lla \z(q_1)\z(-q_1)\rra \lla \z(q_3)\z(-q_3)\rra,
\]
and so overall we recover the leading behaviour \eqref{sqzzzinx}.  These considerations also illustrate the importance of the semi-local contact terms in the numerator of the holographic formula \eqref{zzz}.

\subsubsection{Two scalars and a graviton}

For two scalars and a graviton, in the quasi-equilateral case where all three momenta are of comparable magnitude, inserting our previous results \eqref{OO_FT}, \eqref{TsTs}, \eqref{TsO}, \eqref{Aeq} and \eqref{OOTs_eq_FT}  into the holographic formula \eqref{zzg}, we find
\[\label{zzg_hol}
\lla \z(q_1)\z(q_2)\g^{(s)}(q_3)\rra = \frac{144\,C^2}{\sqrt{2}\pi^2\lambda^4\alpha\alpha_T}\frac{[1+(q_3/q_0)^\lambda]^4}{(q_3/q_0)^{2\lambda}} \frac{\J^2}{a_{123}^2 c_{123}^3 q_3^2} \big[a_{123}^3-a_{123}b_{123}-c_{123}\big],
\]
where $J^2$, $a_{123}$, $b_{123}$ and $c_{123}$ were defined previously in \eqref{sym_polys} and \eqref{J_def}.

\subsubsection{Two gravitons and one scalar}
\label{sec:zgg_hol_discussion}

Turning now to the case of two gravitons and one scalar, 
from our results in Section \ref{sec:stresscorrs} we see that the first term in the numerator of the holographic formula \eqref{zgg} vanishes at leading order in $\lambda$.  Any nonzero contribution must then come from the remaining semi-local terms in the numerator.

The operator $Y_T^{(s_2 s_3)}$ in \eqref{zgg} is defined as the helicity projection
\[
Y_T^{(s_2 s_3)}(q_1)=\frac{1}{4}\ep^{(s_2)}_{ij}(-q_2)\ep^{(s_3)}_{kl}(-q_3)Y^T_{ijkl}(q_1)
\]
of the local operator $Y^T_{ijkl}$ constructed by taking the variation of the stress tensor with respect to the metric,
\[\label{YTdef}
\frac{\delta T_{ij}(x_1)}{\delta g^{kl}(x_2)}\Big|_{g=\delta} \equiv Y^T_{ijkl}(x_1)\delta(x_1-x_2).
\]
From the expression \eqref{Tdef} for the stress tensor, we have
\[
Y^T_{ijkl} = Y^{\T}_{ijkl}+\phi\O \delta_{i(k}\delta_{l)j},
\]
and hence 
\[
Y^{(s_2 s_3)}_T(q_1) = Y^{(s_2 s_3)}_{\T}(q_1)+\frac{1}{4}\phi\O(q_1) \theta^{(s_2 s_3)}
\]
where $\theta^{(s_2 s_3)} = \ep^{(s_2)}_{ij}(-q_2)\ep^{(s_2)}(-q_3)$ and $Y^{\T}_{ijkl}$ is defined as in \eqref{YTdef} replacing $T_{ij}$ with its pure CFT part $\T_{ij}$. (A precise expresion for $\theta^{(s_2s_3)}$ in terms of the momenta may be found in \cite{Bzowski:2011ab}, but we will not need it here.)
The contact term appearing in the numerator of the holographic formulae \eqref{zgg} is then
\[\label{OYT}
\lla \O(q_1)Y_T^{(s_2 s_3)}(-q_1) \rra = \lla \O(q_1)Y_{\T}^{(s_2 s_3)}(-q_1) \rra + \frac{1}{4}\phi\lla \O(q_1)\O(-q_1)\rra \theta^{(s_2s_3)}.
\]
Since the dimension of $Y^{\T}_{ijkl}$ is three, the only way the first correlator on the r.h.s.~could be nonzero would be if an integrated insertion of $\O$ could be contracted with the fixed $Y^{\T}_{ijkl}$ insertion to generate a fixed insertion of $\O$; otherwise we would be left with a CFT 2-point function of two operators with different dimensions which vanishes identically.  
Contractions of this form are governed by the OPE 
\[
\O(x_1) Y^{\T}_{ijkl}(x_2) = D_{ijkl}(x_{12})\O(x_2) + \ldots,
\]
where $D_{ijkl}$ scales as $1/|x_{12}|^3$.  Due to this scaling, there can be no contribution to \eqref{OYT} at order $\lambda^0$ (cf.~our earlier arguments in Section \ref{sec:Tintro} concerning $A_{ij}$); rather, the lowest possible order is $\lambda$.

In fact, we expect one further cancellation to occur for the CFT dual to Einstein gravity: since in slow-roll inflation the correlator $\<\z\g\g\>$ is of order $\lambda^0$ (see \eqref{zggpp_SR} and \eqref{zggpm_SR}), recalling that $\phi=\vphi_1 q_0^\lambda$ is of order $\lambda$ we see the numerator of the holographic formula \eqref{zgg} must be of order $\lambda^2$.    To achieve this, the OPE coefficient $D_{ijkl}$ should have a transverse traceless piece of the right form to produce a cancellation at order $\lambda$ between the correlators on the r.h.s.~of \eqref{OYT}.  
Since $D_{ijkl}$ is a property of the UV CFT it should be computable from AdS/CFT, although we will not pursue this calculation here. 

To obtain the first nonvanishing contribution to $\<\z\g\g\>$ therefore requires an extension of our present analysis to higher order in $\lambda$.  We leave this for future work, but in the meantime note that the inflationary consistency condition for $\<\z\g\g\>$ may be straightforwardly verified through steps parallel to those in Section \ref{sec:3scalarshol}.

Specifically, on dimensional grounds, the function $A(q)$ appearing in the tensor 2-point function \eqref{tensorAdef} must in general be of the form $A(q) = q^3 F(\phi q^{-\lambda})$ for some function $F$, since when $\phi$ vanishes we recover the exact CFT correlator with $A_0(q) \sim q^3$.  On the one hand, then, we have the Callan-Symanzik equation
\[
0=\Big(\frac{\p}{\p\ln q}+\lambda\phi\frac{\p}{\p\phi}-3\Big)\lla T^{(s)}(q)T^{(s')}(-q)\rra,
\]
while on the other hand, from the holographic formula for the tensor power spectrum \eqref{DeltaST}, we have
\[
n_T(q) \equiv \frac{\d \ln \Delta_T^2(q)}{\d\ln q} = 3-\frac{\d \ln A(q)}{\d\ln q}.
\]
Combining these relations we have
\begin{align}\label{OTTnT}
\lambda\phi\lla \O(0)T^{(s)}(q)T^{(s')}(-q)\rra &= -\lambda\phi\frac{\p}{\p\phi}\lla T^{(s)}(q)T^{(s')}(-q)\rra
=\Big(\frac{\p}{\p\ln q} -3\Big)\lla T^{(s)}(q)T^{(s')}(-q)\rra\nn\\&
=-n_T(q)\lla T^{(s)}(q)T^{(s')}(-q)\rra.
\end{align}
Of course in our present leading order analysis these quantities vanish, but in general this will not be so at higher orders.

Considering now the holographic formula \eqref{zgg} for $\<\z\g\g\>$, in the squeezed limit $q_1\ll q_2\approx q_3$ the leading behaviour of the numerator will be governed by the first term, as the remaining semi-local contact terms, which are functions of $q_1$ only, become small.
Approximating this first term using \eqref{OTTnT} above, and using the power spectra formulae \eqref{DeltaST}, we then recover the expected consistency relation \cite{Maldacena:2002vr}
\[
\lim_{q_1\tto 0}\lla \zeta(q_1) \gamma^{(s_2)}(q_2) \gamma^{(s_3)}(q_3) \rra \approx -n_T(q_3) \lla \z(q_1)\z(-q_1)\rra \lla \g^{(s_2)}(q_3)\g^{(s_3)}(-q_3)\rra.
\]

\section{Identifying the dual cosmology} 
\label{sec:Ident}

Our new goal in this section is to identify the slow-roll inflationary background solution to which our perturbed QFT is holographically dual, allowing us to explicitly verify the holographic calculations of the previous section. 

\subsection{Inflationary cosmologies with de Sitter asymptotics} 

Let us consider single-field inflationary cosmologies governed by the action
\[\label{bulkaction}
S= \frac{1}{2\kappa^2}\int \d^4x\sqrt{-g} [R-(\p \Phi)^2 -2\kappa^2 V(\Phi)].
\]
Assuming flat spatial slices, the background FRW solution takes the form 
\[
\d s^2 = -\d t^2 +a^2(t)\d \vec{x}{\,}^2, \qquad \Phi = \vphi(t).
\]
Solutions for which the evolution of the scalar field is monotonic
(or at least piecewise so), as befits an RG flow,
may be obtained from the first-order Hamilton-Jacobi equations of motion \cite{Salopek:1990jq}
\[  \label{eq:bg}
H \equiv \frac{\dot{a}}{a} = -\frac{1}{2}W(\vphi), \qquad \dot{\vphi} = W_{,\vphi},
\]
as may easily be seen by noting that the monotonic function $\vphi(t)$ may be inverted to give $t(\vphi)$ permitting the Hubble rate $H(t)$ to be re-expressed as a function of $\vphi$.  This function, the superpotential $W(\vphi)$, is related to the original potential $V(\vphi)$
by 
\[  
-2\kappa^2V = W_{,\vphi}^2-\frac{3}{2}W^2.
\]
Our interest relates to cosmologies admitting regions in which the metric is asymptotically de Sitter.
Such regions correspond to critical points of the potential, $V_{,\vphi}=0$, for which the superpotential satisfies either
\[
W_{,\vphi}=0, \qquad \mathrm{or} \qquad W_{,\vphi\vphi} = \frac{3}{2} W.
\]
Here we will focus on the first of these possibilities, i.e., solutions for which the critical point of $V(\vphi)$ is also a critical point of $W(\vphi)$.   Cosmologies of this type map under the domain-wall/cosmology correspondence to holographic RG flows that possess `fake' supersymmetry, and such holographic RG flows are known to be stable both perturbatively and non-perturbatively \cite{Freedman:2003ax}.

\begin{figure}[t]
\center
\hspace{-0.6cm}\includegraphics[height=4.5cm]{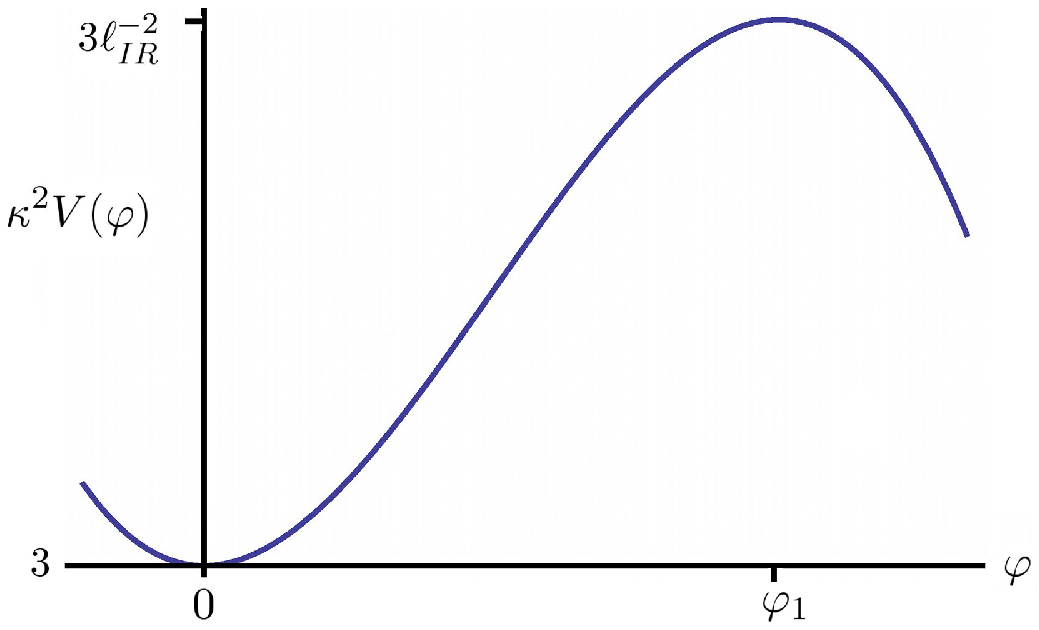}\hspace{0.3cm}  
\includegraphics[height=4.6cm]{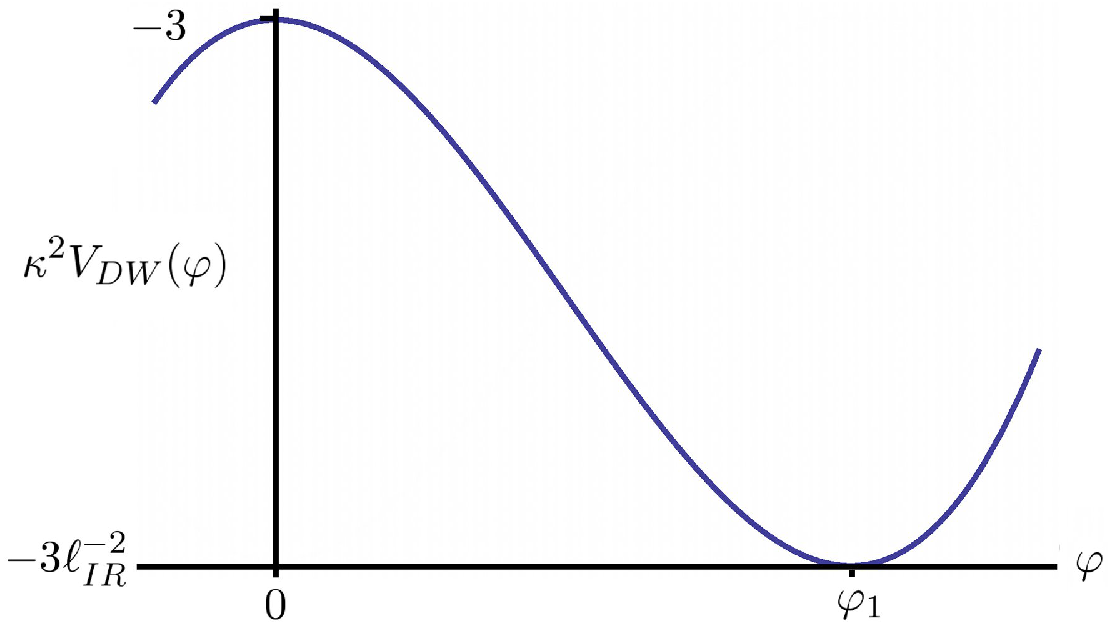}  
\caption{\label{Vplots} The relevant portion of the inflationary potential (left-hand plot) is from the first hilltop at $\phi=\phi_1$ to the origin.  The corresponding domain-wall potential (right-hand plot) is simply minus that of the cosmology.   The factor $\ell_{IR}^{-2}$ is given in \eqref{IRquantities1}. }
\end{figure}

Adopting units in which the de Sitter radius is unity and shifting the scalar field so that the critical point $W_{,\vphi}=0$ occurs at $\vphi=0$, the superpotential may be Taylor expanded about the critical point as
\[
W(\vphi) = -2 +\sum_{n=2}^\infty a_n \vphi^n.
\]
The corresponding potential then takes the form
\[
\kappa^2 V(\vphi) = 3 -a_2(2a_2+3)\vphi^2-3a_3(1+2a_2)\vphi^3+O(\vphi^4).
\]
Under the domain-wall/cosmology correspondence the sign of the potential is reversed (see Fig.~\ref{Vplots}), and hence the potential of the corresponding holographic RG flow is 
\begin{align}
\kappa^2 V_{DW}(\vphi) &= -3 +a_2(2a_2+3)\vphi^2+3a_3(1+2a_2)\vphi^3+O(\vphi^4) \nn\\
& = -3 +\frac{1}{2} m^2\vphi^2 -\frac{1}{3}g\vphi^3+ O(\vphi^4).
\end{align}
The squared mass $m^2 = 2 a_2(2a_3+3)$ and cubic coupling $g=-9a_3(1+2a_2)$ may be related to the operator dimension $\Delta=3-\lambda$ and scalar OPE coefficient $C$ of the dual CFT via the standard AdS/CFT identifications 
\begin{align}
m^2 &=\lambda(\lambda-3) \hspace{0.8cm} \qquad \Rightarrow \qquad \lambda = - 2a_2, \\
g & = -6\pi C (1-\lambda) \qquad \Rightarrow \qquad C = \frac{3}{2\pi} a_3,
\end{align}
as follows from computing holographic correlators for an interacting scalar field on an exact AdS background (see, e.g., \cite{Skenderis:2002wp}).

As our earlier analysis applied to CFTs with $0<\lambda\ll 1$ and $C$ a nonzero positive constant of order unity, we must therefore consider cosmologies with superpotential
\[\label{W}
W(\vphi) = -2 -\frac{1}{2}\lambda\vphi^2 + \frac{2}{3}\pi C\vphi^3 + O(\vphi^4).
\]
From the equation of motion
\[\label{phidotbeta}
\dot{\vphi} = -\lambda \vphi + 2\pi C \vphi^2 +O(\vphi^3),
\]
we see there is then a second critical point of the superpotential close to the origin at
\[
\vphi = \vphi_1 +O(\lambda^2), 
\]
with $\vphi_1 = \lambda/2\pi C$ as before.
Mirroring our earlier analysis of the $\beta$-function, we now specialise to the case of the exact cubic superpotential
by dropping the higher order terms in \eqref{W}, although we expect that our analysis will ultimately hold more generally.\footnote{In particular, since $\vphi$ is small throughout the inflationary evolution, we may perform a field redefinition of the form $\vphi = \bar{\vphi}+O(\bar{\vphi})^3$ to re-introduce higher order terms in the Taylor expansion of the superpotential; note, however, this generates a non-canonical kinetic term in the bulk action \eqref{bulkaction}.
\label{Wn}
A different generalisation would be to consider superpotentials of the form $W(\vphi) =-2-(\lambda/2)\vphi^2+ \tilde{C}_n\vphi^n+O(\vphi^{n+1})$ for which the cubic term is absent, cf.~footnote \ref{betan}.  For these models the background geometry may be solved in the same manner as we discuss here.} 
Integrating \eqref{phidotbeta} with the initial condition $\vphi(t_0)=\vphi_0$ where $0<\vphi_0<\vphi_1$, we obtain
\begin{align}
e^{\lambda (t-t_0)} &= \frac{\vphi_0}{\vphi}\frac{(\vphi_1-\vphi)}{(\vphi_1-\vphi_0)}.
\end{align}
Inverting, we find
\[
\vphi = \vphi_1 \Big[1+\frac{\vphi_1}{\phi}e^{\lambda t}\Big]^{-1}, 
\]
where 
\[
\phi = \vphi_1 e^{\lambda t_0} \Big[\frac{\vphi_1}{\vphi_0}-1\Big]^{-1}
\]
encodes the asymptotic behaviour 
\[
\vphi \tto \phi e^{-\lambda t} \qquad \mathrm{as}\quad t\tto \infty.
\]
From the holographic dictionary, $\phi$ is then the renormalised coupling in the dual QFT, as indeed we would expect from comparison with \eqref{renphi}.
We may also identify the time coordinate $t$ with the RG scale according to $t=-\ln \Lambda$, since $\dot{\vphi}=\beta(\vphi) = -\d\vphi/\d \ln\Lambda$.  

Integrating the equation of motion for the scale factor with boundary condition 
\[
a\tto e^t \qquad \mathrm{as}\qquad t\tto \infty,
\]
we find
\[\label{aofphi}
a^\lambda = \frac{\phi}{\vphi}\Big(1-\frac{\vphi}{\vphi_1}\Big)^{1+\lambda\vphi_1^2/12} \exp\Big[\frac{\lambda}{12}\vphi(\vphi_1-\vphi)\Big],
\] 
which evaluates to
\[
a = \Big(\frac{\phi}{\vphi_1}+e^{\lambda t}\Big)^{-\vphi_1^2/12}
\exp\Big[t\Big(1+\frac{\lambda\vphi_1^2}{12}\Big)+\frac{1}{12}\phi\vphi_1 e^{\lambda t} \Big(\frac{\phi}{\vphi_1}+e^{\lambda t}\Big)^{-2}\,\Big].
\]
In the infinite past, therefore, we obtain the asymptotic behaviour 
\[
a \tto \Big(\frac{\vphi_1}{\phi}\Big)^{\vphi_1^2/12}\exp\Big[t\Big(1+\frac{\lambda\vphi_1^2}{12}\Big)\Big], \qquad \vphi \tto \vphi_1-\frac{\vphi_1^2}{\phi}e^{\lambda t} \qquad \mathrm{as}\qquad t\tto -\infty,
\]
from which we see the difference in Hubble rate between $t\tto-\infty$ and $t\tto \infty$ is $\lambda\vphi_1^2/12 \sim \lambda^3$.
(Interestingly, this difference is proportional to the difference in free energy between UV and IR for the corresponding RG flow on an $S^3$ geometry \cite{Klebanov:2011gs}, $\Delta H = (3/2\pi^4)\Delta F_{\mathrm{sphere}}$.) 

Notice that under an infinitesimal time shift $\delta t = -\sigma$ the background solution for $a$ and $\vphi$ remains invariant provided we simultaneously transform $\delta \phi = -\sigma\lambda\phi$ and $\delta a_0 =\sigma a_0$, where $a_0$ is the boundary scale factor introduced by a rescaling of the spatial coordinates so that $a\tto a_0 e^t$ as $t\tto \infty$.  Thus, minus the time derivative operator in the bulk maps to the dilatation operator $\delta_D = a_0 (\p/\p a_0)-\lambda\phi (\p/\p\phi)$ in the dual QFT.

To determine the parameters $\Delta_{IR}$ and $C_{IR}$ in the IR CFT we examine the Taylor expansion of the potential about $\vphi=\vphi_1$. This takes the form
\[
\kappa^2 V_{DW} = -3 \ell_{IR}^{-2} +\frac{1}{2} m_{IR}^2(\vphi-\vphi_1)^2 -\frac{1}{3}g_{IR}(\vphi-\vphi_1)^3+ O\big((\vphi-\vphi_1)^4\big), 
\]
with
\[\label{IRquantities1}
\ell_{IR}^{-2} = 1+\frac{1}{6}\lambda\vphi_1^2+\frac{1}{144}\lambda^2\vphi_1^2, \quad m_{IR}^2 = 3\lambda+\lambda^2+\frac{1}{4}\lambda^2\vphi_1^2, \quad g_{IR} = -6\pi C(1+\lambda+\frac{1}{12}\lambda\vphi_1^2),
\]
hence from $\Delta_{IR}(\Delta_{IR}-3) = m_{IR}^2 \ell_{IR}^2$ and $g_{IR} = -6\pi C_{IR}(\Delta_{IR}-2)$ we obtain
\[\label{IRquantities2}
\Delta_{IR} =  3 + \lambda - \frac{1}{12}\lambda^2 \vphi_1^2 +O(\lambda^7), \qquad C_{IR} = C +\frac{1}{12}\lambda \vphi_1^2+O(\lambda^4).
\]
Referring back to our earlier discussion below \eqref{betaIR}, the fact that $\Delta_{IR}$ differs from $3+\lambda$ at order $\lambda^4$ suggests that the relation between the bulk inflaton and the dimensionless QFT coupling $\vphi$ may be nontrivial at higher orders in $\lambda$.  Such considerations may be ignored, however, in our present leading order analysis.

\subsection{Slow-roll parameters}  

Let us now evaluate the inflationary power spectra and non-Gaussianities via the slow-roll approximation. 
Introducing the slow-roll parameters
\[
\label{eps_eta_def}
\ep \equiv -\frac{\dot{H}}{H^2} = \frac{2W_{,\vphi}^2}{W^2}, \qquad \eta \equiv \frac{\ddot{\vphi}}{H\dot{\vphi}} = -\frac{2W_{,\vphi\vphi}}{W},
\]
the cosmological power spectra and non-Gaussianities may be expressed in terms of the 
values of these parameters at horizon crossing, defined for some mode of momentum $q$ as the time $t_*$ for which $q=aH$.   These values may be computed as follows.

Expanding \eqref{aofphi}, we find
\[
\vphi_* = \vphi_1\Big[1+\Big(\frac{q}{q_0}\Big)^\lambda\Big]^{-1} + O(\lambda^4),
\]
where $q_0^{-\lambda} = \vphi_1/\phi$ in accordance with \eqref{q0def},
and hence from \eqref{eps_eta_def},
\begin{align}
\label{ep_star}
\ep_* &= \frac{1}{2}\lambda^2\vphi_1^2 \Big(\frac{q}{q_0}\Big)^{2\lambda}\Big[1+\Big(\frac{q}{q_0}\Big)^\lambda\Big]^{-4}+O(\lambda^7), \\[1ex]
\label{eta_star}
\eta_* &= -\lambda+2\lambda\Big[1+\Big(\frac{q}{q_0}\Big)^\lambda\Big]^{-1}+O(\lambda^4), \\[1ex] 
\label{H_star}
H_* & = 1+O(\lambda^3).
\end{align}

\subsection{Power spectra}

Noting that $\ep_* \sim \lambda^4$ while $\eta_* \sim \lambda$, the scalar power spectrum  to leading order in $\lambda$ may be found 
from the usual first-order slow-roll result while the size of the error term may be determined from the second-order slow-roll analysis in \cite{Stewart:1993bc, Gong:2001he}.  This gives    
\begin{align}\label{power_spec1}
\Delta_S^2(q) \equiv \frac{q^3}{2\pi^2}\<\!\<\zeta(q)\zeta(-q)\>\!\> 
&= \frac{\kappa^2 H_*^2}{8\pi^2 \ep_*}+O(\lambda^{-3}) 
= \frac{1}{16}\Delta_S^2(q_0)\Big(\frac{q}{q_0}\Big)^{-2\lambda}\Big[1+\Big(\frac{q}{q_0}\Big)^\lambda\Big]^4 + O(\lambda^{-3}),
\end{align}
where
\[
\Delta_S^2(q_0) = \frac{4\kappa^2}{\pi^2\lambda^2\vphi_1^2} =\frac{16 C^2\kappa^2}{\lambda^4}.
\] 

The spectral tilt
\[\label{spec_tilt2}
n_S-1 = -2\eta_* =2\lambda-4\lambda\Big[1+\Big(\frac{q}{q_0}\Big)^{\lambda}\Big]^{-1}
\]
in striking agreement with our previous holographic result \eqref{spec_tilt}.  
To match the overall amplitude of the power spectrum to our holographic result \eqref{hol_power_spec2} we must normalise the dual scalar operator $\O$ so that to leading order in $\lambda$
\[\label{alpha_norm}
\alpha = \frac{12}{\pi^2\kappa^2},
\]
in agreement with the standard AdS/CFT normalisation of this operator.  
Our holographic result for the scalar power spectrum \eqref{hol_power_spec2} is then in exact agreement with the slow-roll result \eqref{power_spec1}.  Given the non-trivial momentum dependence of these formulae, this agreement is quite remarkable.

The tensor power spectrum may similarly be evaluated from standard slow-roll formulae, yielding
\[
\Delta_T^2(q) = \frac{2 H_*^2}{\pi^2}+O(\lambda^4) = \frac{2}{\pi^2}+O(\lambda^3).
\]
Comparing with our holographic result \eqref{DeltaT_result}, we find exact agreement upon choosing the normalisation of $\T^{(s)}$ to be
\[\label{alpha_T_norm}
\alpha_T = \frac{24}{\pi^2 \kappa^2} = 2\alpha.
\]
This normalisation is again in agreement with the standard AdS/CFT normalisation.
As discussed earlier, to recover the leading nonzero contribution to the tensor tilt $n_T \sim \ep_* \sim \lambda^4$ would require an extension of our holographic analysis to higher order in $\lambda$.

\subsection{Non-Gaussianities} 
\label{sec:nGcosmo}

The cosmological 3-point functions for scalar and tensor fluctuations were evaluated for quasi-equilateral and for squeezed momentum configurations at leading order in slow-roll in  \cite{Maldacena:2002vr}.\footnote{At leading order, the slow-roll parameters $\ep_V$ and $\eta_V$ used in \cite{Maldacena:2002vr} are related to those here by $\ep_V=\ep$, $\eta_V=\ep-\eta$.}   The various contractions of helicity tensors appearing in the formulae of \cite{Maldacena:2002vr} may be re-expressed in terms of the momenta and helicities using the formulae presented in Appendix C of \cite{McFadden:2011kk} (see also Appendix A of \cite{Bzowski:2011ab}).  After this step, the slow-roll 3-point functions reduce to completely explicit functions of the momenta.  

\subsubsection{Three scalars}
\label{sec:nGcosmozzz}

In the quasi-equilateral case \eqref{quasi_eq} where all the $q_i^{-\lambda}$ are of comparable magnitude, we have
\begin{align}
\label{zzz_SR}
\<\!\<\z(q_1)\z(q_2)\z(q_3)\>\!\> & = \frac{\kappa^4 H_*^4 \eta_*}{16 \ep_*^2}\sum_{i<j}q_i^{-3} q_j^{-3} + O(\lambda^{-4}).
\end{align}
Here, the leading term is of order $H_*^4\eta_*/\ep_*^2\sim \lambda^{-7}$ while all remaining terms in the slow-roll result of \cite{Maldacena:2002vr} for this correlator are proportional to $H_*^4/\ep_*\sim \lambda^{-4}$ and thus are subleading in the $\lambda$ expansion.  
The leading order scalar bispectrum is therefore of the local type with $f_{NL}=5\eta_*/6$.

Inserting our results \eqref{ep_star}--\eqref{H_star} for $\ep_*$, $\eta_*$ and $H_*$ with $q_3$ as our reference momentum, 
and making use of the normalisation \eqref{alpha_norm}, we may rewrite \eqref{zzz_SR} as 
\[
\<\!\<\z(q_1)\z(q_2)\z(q_3)\>\!\> = 
\frac{576\, C^4}{\alpha^2 \lambda^7}
\Big({-}1+2\Big[1+\Big(\frac{q_3}{q_0}\Big)^{\lambda}\Big]^{-1}\Big)
\Big(\frac{q_3}{q_0}\Big)^{-4\lambda}
\Big[1+\Big(\frac{q_3}{q_0}\Big)^{\lambda}\Big]^8 \sum_{i<j}q_i^{-3}q_j^{-3} +O(\lambda^{-7}).
\]
Comparing with our holographic result \eqref{zzz_hol_eq} for the corresponding quasi-equilateral case, we find perfect agreement.
Moreover, as we have already verified in Section \ref{sec:3scalarshol}, the behaviour of the holographic result in the squeezed limit
(cf.~\eqref{sqzzzinx}) is in agreement with that given by the consistency relation of \cite{Maldacena:2002vr}.

\subsubsection{Two scalars and one graviton}

In the quasi-equilateral case \eqref{quasi_eq}, the complete result for this correlator at first order in slow-roll is
\begin{align}
\label{zzg_SR}
\<\!\<\z(q_1)\z(q_2)\g^{(\pm)}(q_3)\>\!\> & = \frac{\K^4 H_*^4}{16\sqrt{2}\ep_*} \frac{\J^2}{a_{123}^2 c_{123}^3 q_3^2} \big[a_{123}^3-a_{123}b_{123}-c_{123}\big],
\end{align}
where the elementary symmetric polynomials $a_{123}$, $b_{123}$ and $c_{123}$ were given in \eqref{sym_polys} and $J^2$ was defined in \eqref{J_def}.

Inserting our results \eqref{ep_star} and \eqref{H_star} for $\ep_*$ and $H_*$ with $q_3$ as our reference momentum, 
and using the normalisations \eqref{alpha_norm} and \eqref{alpha_T_norm}, we find 
\[\label{zzg_SR2}
\lla \z(q_1)\z(q_2)\g^{(s)}(q_3)\rra = \frac{144\,C^2}{\sqrt{2}\pi^2\lambda^4\alpha\alpha_T}\frac{[1+(q_3/q_0)^\lambda]^4}{(q_3/q_0)^{2\lambda}} \frac{\J^2}{a_{123}^2 c_{123}^3 q_3^2} \big[a_{123}^3-a_{123}b_{123}-c_{123}\big].
\]
This result matches exactly our holographic expression \eqref{zzg_hol}.  Once again, this agreement is striking in view of the nontrivial momentum dependence of this formula.

\subsubsection{Two gravitons and one scalar}

For completeness, the first-order slow-roll correlator for two gravitons and a scalar is given in the quasi-equilateral case \eqref{quasi_eq} by
\begin{align}
\label{zggpp_SR}
\<\!\<\z(q_1)\g^{(\pm)}(q_2)\g^{(\pm)}(q_3)\>\!\> &= -\frac{\K^4H_*^4}{128 b_{23}^5 q_1^2}(q_1^2-a_{23}^2)^2
\Big[(q_1^2-a_{23}^2+2b_{23})-\frac{8 b_{23}^2}{q_1 a_{123}}\Big],
\\[2ex]
\label{zggpm_SR}
\<\!\<\z(q_1)\g^{(\pm)}(q_2)\g^{(\mp)}(q_3)\>\!\> &= -\frac{\K^4H_*^4}{128\, b_{23}^5 q_1^2}(q_1^2-a_{23}^2+4b_{23})^2\big[(q_1^2-a_{23}^2+2b_{23})-\frac{8 b_{23}^2}{q_1 a_{123}}\big],
\end{align}
where we have supplemented the elementary symmetric polynomials in \eqref{sym_polys} with $a_{23} = q_2+q_3$ and $b_{23} = q_2q_3$.
As we discussed in Section \ref{sec:zgg_hol_discussion}, however, to recover this first nonzero contribution to $\<\z\g\g\>$ holographically requires working to higher order in $\lambda$.  
Such an analysis should also be able to recover the subleading equilateral pieces of order $\lambda^{-4}$ in $\<\z\z\z\>$.

\section{Discussion}
\label{sec:disc}

In this work we have constructed a holographic duality between a single-field inflationary slow-roll cosmology and a three-dimensional QFT consisting of a CFT plus a single nearly marginal scalar deformation.  Our methods enable explicit computations to be performed on both sides of the duality allowing our holographic framework to be directly verified.  The form of inflationary correlators is seen to be controlled by the perturbative breaking of conformal symmetry in the vicinity of a fixed point of the dual QFT,
completely determining the spectra and bispectra up to a few constants.  If these constants are fixed using AdS/CFT at the fixed point we obtain a model corresponding to Einstein gravity, otherwise we obtain alternative models.

It should be emphasised that our conformal perturbation expansion in the anomalous dimension $\lambda\ll 1$ 
leads to slow-roll parameters $\ep_*\sim \lambda^4$ while $\eta_*\sim \lambda$.   Working at leading order in $\lambda$, we have been able to recover the slow-roll $\<\z\z\>$ and $\<\z\z\g\>$ correlators exactly, along with the leading order pieces of all other 
correlators, $\<\g\g\>$,  $\<\z\z\z\>$ and $\<\z\g\g\>$ (although this last correlator is zero to leading order).  To
obtain the remaining pieces, such as the tensor tilt or the leading non-zero part of the $\<\z\g\g\>$ correlator will require further calculations at higher order in $\lambda$.
Alternatively, one might look for modifications of the present set-up in which $\ep_*$ and $\eta_*$ are both of the same order.

In this paper we were able to recover the entire momentum dependence of the cosmological correlators by resumming an infinite 
set of diagrams in conformal perturbation theory. Naively, in conformal perturbation theory the leading order contribution 
to the 2-point (3-point) functions comes from the integrated  3-point (4-point) functions. This expectation turned out to be incorrect in our case. When the dimension of the deforming scalar operator is close to three, $\Delta = 3-\lambda$,
the integrated higher point functions are singular in $\lambda$ and one needs to resum an infinite number of insertions.
The coefficient of the leading order singularity, however, which is the quantity entering our leading order computations, 
is universal.  This appears to be related to a (new?)~conformal anomaly present when there are marginal operators.
Correlation functions of such operators contain logarithmic terms, and these are linked with the singular terms in the correlators of the $\Delta = 3-\lambda$ operator. We hope to discuss this issue in more detail elsewhere.

The naive application of conformal perturbation theory still applies if in addition we focus on the leading short/long-wavelength 
behaviour of the correlators, since the higher point functions give subleading contributions in this limit.
Focusing on such terms only,
one may then analyse the subleading in $\lambda$ contributions to the correlators.  Using this approach we were also able 
to recover the subleading contribution  in slow-roll to the scalar power spectrum \cite{Gong:2001he},
\[
\Delta_S^2(q) = \frac{H_*^2}{(2\pi)^2 2\ep_*}\left[1+2 ( 2-\ln 2 -\gamma) \eta_*\right]  + O(\lambda^{-2}).
\]
It would be interesting to extend this computation to the other observables as well.

Our approach may readily be extended to other bulk actions beyond standard Einstein gravity.  For example, 
it would be interesting to consider actions allowing for a non-zero CFT 3-point function $\<\T_{ij}(x_1)\T_{kl}(x_2)\O(x_3)\>_0$.\footnote{
It seems likely that the action (\ref{bulkaction}) with  the additional term $\Phi C_{ijkl} C^{ijkl}$, where $C_{ijkl}$ is the Weyl tensor, would have this property.} The nature of the resumming involved to calculate correlators in the perturbed QFT would be modified leading to different holographic predictions. 
We should  also emphasise that the results from conformal perturbation theory are valid irrespectively of whether the CFT is at strong or at weak coupling. A weakly coupled CFT would correspond to a strongly coupled non-geometric bulk and we may extract from our results the phenomenology of such models.  

In this paper we found holographically that all leading order results are essentially fixed by the broken conformal invariance.
It would be interesting to analyse the same question directly from the bulk perspective. 

The holographic construction leads to a connection between properties of CFTs and inflationary physics. In particular, single scalar models are linked to CFTs that contain in their spectrum a single scalar operator which is nearly marginal. The presence of additional nearly marginal operators would map to the presence of bulk light scalars, and the inflationary model would then be a multi-scalar one. 
In general, the deforming operator would mix with other operators along the flow corresponding to the effects of entropy perturbations in the cosmology. 
It would be very interesting to map out all such possibilities and classify inflationary models using 
properties of the dual QFT. It would also be very interesting to find concrete CFT models with the required spectrum and OPE structure. We intend to pursue these and related questions in future work.


\bigskip

{\it Acknowledgments:}  
Research at the Perimeter Institute is supported by the Government of Canada
through Industry Canada and by the Province of Ontario through the Ministry of
Research \& Innovation. K.S.~acknowledges support from NWO via a VICI grant and from a grant of the John Templeton Foundation.  The opinions expressed in this publication are those of the authors and do not necessarily reflect the views of the John Templeton Foundation. A.B.~is supported through NWO via the VICI grant of K.S.

\appendix

\section{Fourier transform of correlators}

In this appendix we evaluate the Fourier transforms of the various 3-point correlators in the main text, both for the exact CFT and for the perturbed theory.  

\subsection{Scalar 3-point function}
\label{app:OOO_FT}

Let us begin by evaluating the Fourier transform of an exact CFT correlator with three arbitrary scalar insertions,
\[
\< \O_{\Delta_1}(x_1)\O_{\Delta_2}(x_2)\O_{\Delta_3}(x_3)\>_0 = \frac{\alpha C}{|x_{12}|^{\beta_3}|x_{23}|^{\beta_1}|x_{31}|^{\beta_2}},
\]
where $\beta_i = \Delta_t-2\Delta_i$ and $\Delta_t = \sum_i \Delta_i$.  Dropping the overall delta function from momentum conservation, we have
\[
\lla \O_{\Delta_1}(q_1)\O_{\Delta_2}(q_2)\O_{\Delta_3}(q_3)\rra_0 = \alpha C \I(\beta_1,\beta_2\,\beta_3),
\]
where
\[\label{Ibetadef}
\I(\beta_1,\beta_2\,\beta_3) = \int\d^3 x_{31}\int\d^3 x_{23} \frac{e^{iq_1\cdot x_{31}-i q_2\cdot x_{23}}}{|x_{23}|^{\beta_1}|x_{31}|^{\beta_2}|x_{23}+x_{31}|^{\beta_3}}.
\]
Using the Schwinger representation
\[
\frac{1}{|x|^\beta} = \frac{1}{\Gamma(\beta/2)}\int_0^\infty \d^3t \, t^{\frac{\beta}{2}-1}e^{-tx^2}
\]
to perform the integrals over $x_{23}$ and $x_{31}$, we obtain
\[
\I(\beta_1,\beta_2,\beta_3) = \int_{\mathbb{R}^3_+}\d^3 \vec{t}\, \frac{\pi^3}{T^{3/2}}\prod_i \frac{1}{\Gamma(\beta_i/2)} t_i^{\frac{\beta_i}{2}-1} \exp\Big({-}\frac{t_i q_i^2}{4T}\Big),
\]
where $T = t_1t_2+t_2t_3+t_3t_1$.  After a change of variables
\[
u_i = \frac{t_i q_i^2}{4T},\qquad U = \frac{1}{T}, \qquad |\d t_1\wedge \d t_2\wedge \d t_3| = \frac{64}{U^3 q_1^2q_2^2q_3^2}|\d u_1\wedge \d u_2\wedge \d u_3|,
\]
we find
\[
\I(\beta_1,\beta_2,\beta_3) = \int_{\mathbb{R}^3_+} \d^3\vec{u}\, U^{(3-\beta_t)/2}\prod_i \frac{\pi}{\Gamma(\beta_i/2)}\Big(\frac{4u_i}{q_i^2}\Big)^{\beta_i/2} u_i^{-1}e^{-u_i},
\]
where $\beta_t=\sum_i\beta_i$.
Exponentiating the factor of $U$ with a further Schwinger parameter according to
\[
U^{(3-\beta_t)/2} = \frac{2}{\Gamma((\beta_t-3)/2)}\int_0^\infty \d\rho \,\rho^{\beta_t-4} \prod_i \Big(\frac{4u_i}{q_i^2}\Big)^{(3-\beta_t)/2} \exp\Big({-}\frac{\rho^2q_i^2}{4u_i}\Big), 
\]
we then use a standard representation for modified Bessel functions of the second kind,
\[
K_{\nu}(\rho q) = \frac{1}{2}\Big(\frac{\rho q}{2}\Big)^\nu \int_0^\infty \d u\, u^{-\nu-1} \exp\Big({-}u-\frac{\rho^2q^2}{4u}\Big),
\]
to rewrite the integral as
\[\label{tripleKresult}
\I(\beta_1,\beta_2,\beta_3) = \frac{\pi^3\,2^{17/2-\beta_t}}{\Gamma((\beta_t-3)/2)}\int_0^\infty \d\rho\sqrt{\rho}\prod_i\frac{1}{\Gamma(\beta_i/2)}q_i^{v_i}K_{\nu_i}(\rho q_i),
\]
where 
\[
\nu_i = \frac{1}{2}(\beta_t-\beta_i-3).
\]
The result \eqref{tripleKresult} may be thought of as a Witten diagram with three AdS bulk-to-boundary propagators meeting at a bulk vertex.
Formally this integral is divergent, since 
\[
K_\nu(\rho q) \approx \frac{1}{2}\Gamma(\nu) \Big(\frac{\rho q}{2}\Big)^{-\nu} \qquad \mathrm{as}\qquad \rho \tto 0,
\]
reflecting the fact that in taking the Fourier transform we integrated over configurations in which the insertion points were coincident.  
This unphysical divergence may be removed in a simple fashion\footnote{A more sophisticated option might be to apply differential regularisation \cite{Freedman:1991tk}.} through analytic continuation, noting that the integral in \eqref{tripleKresult} converges when the power of $\rho$ in the integrand is increased.
The result may be expressed in terms of Appell $F_4$ \cite{Prudnikov,Boos:1987bg, Anastasiou:1999ui} in the form
\begin{align}\label{F4result}
& \int_0^\infty \d \rho \,\rho^{\sigma+1/2} K_{\nu_1}(\rho q_1)
K_{\nu_2}(\rho q_2)K_{\nu_3}(\rho q_3) \nn\\&\qquad
= 2^{-5/2+\sigma}q_3^{-3/2-\sigma}[A(\nu_1,\nu_2)+A(-\nu_1,\nu_2)+A(\nu_1,-\nu_2)+A(-\nu_1,-\nu_2)],
\end{align}
where
\begin{align}
&A(\nu_1,\nu_2) = \nn\\
&\qquad \Big(\frac{q_1}{q_3}\Big)^{\nu_1}\Big(\frac{q_2}{q_3}\Big)^{\nu_2} \Gamma\Big(\frac{1}{2}\Big(\frac{3}{2}+\sigma+\nu_1+\nu_2-\nu_3\Big)\Big)
\Gamma\Big(\frac{1}{2}\Big(\frac{3}{2}+\sigma+\nu_1+\nu_2+\nu_3\Big)\Big)
\Gamma(-\nu_1)\Gamma(-\nu_2)\nn\\&\qquad\times F_4\Big[\frac{1}{2}\Big(\frac{3}{2}+\sigma+\nu_1+\nu_2-\nu_3\Big),\frac{1}{2}\Big(\frac{3}{2}+\sigma+\nu_1+\nu_2+\nu_3\Big);
\nu_1+1,\nu_2+1; \frac{q_1^2}{q_3^2}, \frac{q_2^2}{q_3^2}\Big].
\end{align}
For general arguments, the domain of convergence of this result is 
\[
\mathrm{Re}\, \sigma > \sum_i|\mathrm{Re}\, \nu_i|-\frac{3}{2}, \qquad \mathrm{Re}\,\Big(\sum_i q_i\Big) >0.
\]
Appell $F_4$ may itself be defined by analytic continuation of the double hypergeometric series 
\[\label{F4def}
F_4[a,b;c,d; x,y] = \sum_{m,n=0}^\infty \frac{(a)_{m+n}(b)_{m+n}}{(c)_m (d)_n m! n!}x^m y^n, \qquad \sqrt{|x|}+\sqrt{|y|}<1,
\]
where the Pochammer symbol $(a)_m = \Gamma(a+m)/\Gamma(a)$.  
In particular, the continuation of \eqref{F4def} to triangle momentum configurations is given in \cite{Exton, Alkofer:2008dt}. 
In the following, however, we will fortunately be able to evaluate the specific cases arising by simpler means.

Let us now consider the scalar 3-point function for the perturbed QFT.  As we saw in \eqref{OOO_as_sum}, this may be expressed as a sum of CFT correlators with shifted dimensions.  Making use of \eqref{tripleKresult}, this sum takes the form
\begin{align}\label{tripleK}
\<\!\<\O(q_1)\O(q_2)\O(q_3)\>\!\> & =   {\sum_{\ell_1,\ell_2,\ell_3=0}^{\infty}} B(\ell_1,\ell_2,\ell_3)
\int_0^\infty \d \rho \, \sqrt{\rho} K_{\mu_1}(\rho q_1)
K_{\mu_2}(\rho q_2)K_{\mu_3}(\rho q_3),
\end{align}
where
\begin{align}\label{nu_defs}
\mu_i &= \frac{3}{2}-\frac{\lambda}{2}(\ell_t-\ell_i+2), \\[1ex]
B(\ell_1,\ell_2,\ell_3) &= \frac{\pi^3\alpha\,C}{\sqrt{2}}  \frac{2^{(\ell_t+3)\lambda}}{\Gamma\big(3-(\ell_t+3)\lambda/2\big)}
\Big({-}\frac{\phi}{\vphi_1}\Big)^{\ell_t} \prod_i \frac{q_i^{\mu_i}(\ell_i+1)}{\Gamma\big((3-(\ell_i+1)\lambda)/2\big)} \nn\\
&= (2\pi)^{3/2}\alpha\,C \prod_i q_i^{\mu_i}\Big({-}\frac{\phi}{\vphi_1}\Big)^{\ell_i}(\ell_i+1) + O(\lambda).
\end{align}
To evaluate these triple-Bessel integrals via \eqref{F4result}, we only need the results 
\begin{align}
F_4\Big[\frac{\lambda}{2}(\ell_2+1),\frac{3}{2};-\frac{1}{2},\frac{5}{2};\frac{q_1^2}{q_3^2},\frac{q_2^2}{q_3^2}\Big] &= 1+O(\lambda), \nn\\
F_4\Big[\frac{\lambda}{2}(\ell_3+1),-\frac{3}{2};-\frac{1}{2},-\frac{1}{2};\frac{q_1^2}{q_3^2},\frac{q_2^2}{q_3^2}\Big] &= 1+O(\lambda),
\end{align}
which follow directly from the series definition \eqref{F4def} choosing $q_3$ to be the largest momentum so that the arguments of 
the Appell function are bounded by unity.
Inserting the values of $\mu_i$ in \eqref{nu_defs}, we find that \eqref{F4result} converges for $\sigma=0$ providing a unique definition for the triple-Bessel integral. At leading order in $\lambda$, this is
\begin{align}\label{tripleKredux}
\int_0^\infty \d \rho \,\sqrt{\rho} K_{\mu_1}(\rho q_1)K_{\mu_2}(\rho q_2)K_{\mu_3}(\rho q_3)  &= \frac{1}{3\lambda}\Big(\frac{\pi}{2}\Big)^{3/2}q_3^{-3/2} \Big[ \frac{1}{(\ell_1+1)}\Big(\frac{q_1}{q_3}\Big)^{\mu_1}\Big(\frac{q_2}{q_3}\Big)^{-\mu_2} \nn\\&\quad
{+} \frac{1}{(\ell_2+1)}\Big(\frac{q_1}{q_3}\Big)^{-\mu_1}\Big(\frac{q_2}{q_3}\Big)^{\mu_2}
{+} \frac{1}{(\ell_3+1)}\Big(\frac{q_1}{q_3}\Big)^{-\mu_1}\Big(\frac{q_2}{q_3}\Big)^{-\mu_2}\Big].
\end{align}
Plugging \eqref{tripleKredux} back into \eqref{tripleK} and resumming the binomial series over the $\ell_i$, we arrive at the result \eqref{OOO_FT} for the Fourier transform quoted in the main text.

\subsection{3-point function of the stress tensor and two scalars}
\label{app:OOT_FT}

We now evaluate the Fourier transform of the exact CFT 3-point function of two scalars and a stress tensor given in \eqref{CFT_OOT}.  Once again, we will use this result as a stepping stone to evaluate the corresponding correlator \eqref{sum_formula_OOT} in the perturbed theory.  

From \eqref{CFT_OOT}, we have the Fourier transform
\begin{align}
\<\!\<\O_{\Delta}(q_1)\O_{\Delta}(q_2)\T_{ij}(q_3)\>\!\>_0
= \int\d^3 x_{31}\int\d^3x_{23} \frac{e^{iq_1\cdot x_{31}-iq_2\cdot x_{23}}}{|x_{23}+x_{31}|^{3-2\lambda}|x_{23}|^3 |x_{31}|^3}\,t_{ij}(X), 
\end{align}
where, expanding the leading order expression \eqref{tX_def},
\[\label{tX_expanded}
t_{ij}(X) = \frac{3\alpha}{8\pi}\delta_{ij}- \frac{9 \alpha}{8\pi}\frac{ x_{31}^2 x_{23}^2}{(x_{23}+x_{31})^2}\Big(\frac{x_{31i}x_{31j}}{x_{31}^4}+\frac{2x_{31(i}x_{23j)}}{x_{31}^2x_{23}^2}+\frac{x_{23i}x_{23j}}{x_{23}^4}\Big).
\]
In terms of the integral $\I(\beta_1,\beta_2,\beta_3)$ defined in \eqref{Ibetadef}, we may rewrite this as
\begin{align}
&\<\!\<\O_{\Delta}(q_1)\O_{\Delta}(q_2)\T_{ij}(q_3)\>\!\>_0 = \nn\\&\qquad
\frac{9\alpha}{8\pi}\Big[\delta_{ij}\Big(\frac{1}{3}\I(3,3,3{-}2\lambda){+}\frac{1}{q_1}\frac{\p}{\p q_1}\I(1,5,5{-}2\lambda){+}\frac{1}{q_2}\frac{\p}{\p q_2}\I(5,1,5{-}2\lambda)\Big) \nn\\& \qquad\qquad
+q_{1i}q_{1j}\frac{1}{q_1}\frac{\p}{\p q_1}\Big(\frac{1}{q_1}\frac{\p}{\p q_1}\I(1,5,5{-}2\lambda)\Big)
-2q_{1(i}q_{2j)}\frac{1}{q_1q_2}\frac{\p}{\p q_1}\frac{\p}{\p q_2}\I(3,3,5{-}2\lambda) \nn\\&\qquad\qquad
+q_{2i}q_{2j}\frac{1}{q_2}\frac{\p}{\p q_2}\Big(\frac{1}{q_2}\frac{\p}{\p q_2}\I(5,1,5{-}2\lambda)\Big)\Big].
\end{align}
In fact, as we will eventually migrate to a helicity basis, we are really only interested in the transverse traceless piece of this result obtained by projecting with $\Pi_{ijkl}(q_3)$ (given in \eqref{proj_op_x}).  Since momentum conservation $\sum_i q_i=0$ implies $\Pi_{ijkl}(q_3)q_{1k}q_{1l}=-\Pi_{ijkl}(q_3)q_{1(k}q_{2l)}=\Pi_{ijkl}(q_3)q_{2k}q_{2l}$, we have
\[
\<\!\<\O_{\Delta}(q_1)\O_{\Delta}(q_2)\T^\perp_{ij}(q_3)\>\!\>_0 = \A_0(q_1,q_2,q_3) \Pi_{ijkl}(q_3)q_{1k}q_{1l},
\]
where
\begin{align}
\A_0(q_1,q_2,q_3) &= 
\frac{9\alpha}{8\pi}\Big[
\frac{1}{q_1}\frac{\p}{\p q_1}\Big(\frac{1}{q_1}\frac{\p}{\p q_1}\I(1,5,5{-}2\lambda)\Big)
+\frac{2}{q_1q_2}\frac{\p}{\p q_1}\frac{\p}{\p q_2}\I(3,3,5{-}2\lambda) \nn\\&\qquad\qquad
+\frac{1}{q_2}\frac{\p}{\p q_2}\Big(\frac{1}{q_2}\frac{\p}{\p q_2}\I(5,1,5{-}2\lambda)\Big)\Big].
\end{align}
Noting that
\[
\frac{1}{q}\frac{\p}{\p q}\Big(q^\nu K_\nu(\rho q)\Big) = -\rho q^{\nu-1} K_{\nu-1}(\rho q),
\]
we may use \eqref{tripleK} to obtain
\[\label{TTO_tripleK}
\A_0(q_1,q_2,q_3) = \frac{\alpha}{3}\sqrt{2\pi}\, q_1^{3/2-\lambda}q_2^{3/2-\lambda}q_3^{3/2}\int_0^\infty \d\rho\,\rho^{5/2} K_{3/2-\lambda}(\rho q_1)K_{3/2-\lambda}(\rho q_2)K_{3/2}(\rho q_3).
\]

In the quasi-equilateral case \eqref{quasi_eq}, this integral is equivalent at leading order to
\[
\A_{eq}(q_1,q_2,q_3) = \frac{\alpha}{3}\sqrt{2\pi}\, (q_1 q_2 q_3)^{3/2} q_3^{-2\lambda}\int_0^\infty \d\rho\,\rho^{5/2} K_{3/2}(\rho q_1)K_{3/2}(\rho q_2)K_{3/2}(\rho q_3),
\]
as may be confirmed from \eqref{F4result}.
This integral may then be evaluated by elementary means after analytically continuing the power of $\rho$ in the integrand to larger values, yielding the simple result
\[\label{A0eq}
\A_0^{eq}(q_1,q_2,q_3) = \frac{\alpha\pi^2}{6} q_3^{-2\lambda}\frac{(-a_{123}^3+a_{123}b_{123}+c_{123})}{a_{123}^2},
\]
where the elementary symmetric polynomials appearing in this formula are given in \eqref{sym_polys}.

To evaluate the triple-Bessel function integral \eqref{TTO_tripleK} for general momenta is considerably more involved, but once again may be accomplished with the help of \eqref{F4result}.  
This time, the Appell $F_4$ functions may be reduced to ordinary Gauss ${}_2F_1$ hypergeometric functions via standard reduction relations \cite{Bateman}.  Although we will not actually use these results in the main text, we provide them here for completeness.
In manipulating the resulting expressions, it is useful to introduce the triangle angles 
\begin{align}
\cos \theta_1 = \frac{-q_1^2+q_2^2+q_3^2}{2q_2q_3}, \qquad \sin \theta_1= \frac{\J}{2q_2q_3},
\end{align}
where $\J^2$ is given in \eqref{J_def} and analogous definitions for $\theta_2$ and $\theta_3$ follow by cyclic permutation.  
Thus, if we take $q_3$ to be our reference momentum, defining
\[
X=\frac{q_1}{q_2}e^{i\theta_3}, \qquad Y = \frac{q_2}{q_1}e^{i\theta_3},
\]
we have, e.g.,
\begin{align}
F_4\Big[\frac{5}{2},4;\frac{5}{2},\frac{5}{2};\frac{q_1^2}{q_3^2},\frac{q_2^2}{q_3^2}\Big] &=
F_4\Big[\frac{5}{2},4;\frac{5}{2},\frac{5}{2};\frac{-X}{(1-X)(1-Y)},\frac{-Y}{(1-X)(1-Y)}\Big] \nn\\
&= (1-X)^4(1-Y)^4\,{}_2F_1\Big[4,\frac{5}{2},\frac{5}{2};XY] \nn\\
&= \frac{(1-X)^4(1-Y)^4}{(1-XY)^4} = \frac{q_3^8}{\J^4}.
\end{align}
The other Appell functions arising may be similarly dealt with:
\begin{align}
F_4\Big[1,\frac{5}{2};\frac{5}{2},-\frac{1}{2}; \frac{q_1^2}{q_3^2},\frac{q_2^2}{q_3^2}\Big] &=
(1-X) \,{}_2F_1\Big[1,\frac{5}{2},-\frac{1}{2};\frac{-Y(1-X)}{(1-Y)}\Big] \nn\\
&=\Big(1-\frac{q_1}{q_2}e^{i\theta_3}\Big) \,{}_2F_1\Big[1,\frac{5}{2},-\frac{1}{2};e^{-2i\theta_1}\Big]\nn\\
&= -\J^{-4}q_3^2(-q_1^2+q_2^2+q_3^2)(\J^2+8q_2^2q_3^2), 
\end{align}
\begin{align}
F_4\Big[-\frac{1}{2},1;-\frac{1}{2},-\frac{1}{2}; \frac{q_1^2}{q_3^2},\frac{q_2^2}{q_3^2}\Big] &=
(1-X)(1-Y)\,{}_2F_1\Big[1,\frac{5}{2},-\frac{1}{2};XY\Big] \nn\\
&=\frac{(1-X)(1-Y)}{(1-XY)^4}\,(1-9XY-9X^2Y^2+X^3 Y^3) \nn\\
&= \J^{-4}q_3^2(q_1^2+q_2^2-q_3^2)(\J^2+8q_1^2q_2^2).
\end{align}

The final result, to leading order in $\lambda$, takes the form
\begin{align}\label{A0_result}
\A_0(q_1,q_2;q_3) &= \frac{\alpha\pi^2}{6}\frac{1}{\J^4}\Big[16 q_1^{3-2\lambda}q_2^{3-2\lambda}q_3^{3+2\lambda} 
+q_1^{3-2\lambda}(q_1^2-q_2^2-q_3^2)(\J^2+8q_2^2q_3^2) \nn\\
&\quad +q_2^{3-2\lambda}(q_2^2-q_1^2-q_3^2)(\J^2+8q_1^2q_3^2) 
+q_3^{3-2\lambda}(q_3^2-q_1^2-q_2^2)(\J^2+8q_1^2q_2^2)\Big],
\end{align}
where we have used a semicolon in the argument on the l.h.s.~to indicate that $q_3$ is our (implicitly nonzero) reference momentum.  

In cases where we cannot assume $q_3$ to be nonzero, we may instead take $q_1$ as our nonzero reference momenta, choosing $q_1$ to be the largest of the two momenta associated with scalar insertions.  In this case, we obtain instead
\begin{align}\label{other_A0_result}
\A_0(q_1;q_2,q_3) &= \frac{\alpha\pi^2}{6}\frac{1}{\J^4}\Big[16 q_1^{3}q_2^{3-2\lambda}q_3^{3} 
+q_1^{3-2\lambda}(q_1^2-q_2^2-q_3^2)(\J^2+8q_2^2q_3^2) \nn\\
&\quad +q_2^{3-2\lambda}(q_2^2-q_1^2-q_3^2)(\J^2+8q_1^2q_3^2) 
+q_1^{-2\lambda}q_3^3(q_3^2-q_1^2-q_2^2)(\J^2+8q_1^2q_2^2)\Big].
\end{align}
In the quasi-equilateral limit \eqref{quasi_eq} when all three momenta are comparable,
both results \eqref{A0_result} and \eqref{other_A0_result} reduce to \eqref{A0eq} above.

To evaluate the correlator $\<\!\<\O(x_1)\O(x_2)\T^\perp_{ij}(x_3)\>\!\>$ in the perturbed theory, we use the result \eqref{sum_formula_OOT} expressing this correlator as a sum of CFT correlators with shifted dimensions.
Shifting the argument $\lambda \tto (1+n/2)\lambda$ in \eqref{A0_result}, inserting into \eqref{sum_formula_OOT} and resumming the binomial series, we find
\[
\<\!\<\O_{\Delta}(q_1)\O_{\Delta}(q_2)\T^\perp_{ij}(q_3)\>\!\> = \A(q_1,q_2,q_3) \Pi_{ijkl}(q_3)q_{1k}q_{1l},
\]
where for nonzero $q_3$,
\begin{align}
\A(q_1,q_2;q_3) &= \frac{\alpha\pi^2}{6}\frac{1}{\J^4}
\Big[16 q_1^{3-2\lambda}q_2^{3-2\lambda}q_3^{3+2\lambda}\Big[1+\frac{\phi}{\vphi_1}\Big(\frac{q_1q_2}{q_3}\Big)^{-\lambda}\Big]^{-4}  \nn\\&\qquad
+\Big[1+\frac{\phi}{\vphi_1}q_1^{-\lambda}\Big]^{-4}q_1^{3-2\lambda}(q_1^2-q_2^2-q_3^2)(\J^2+8q_2^2q_3^2) \nn\\
&\qquad +\Big[1+\frac{\phi}{\vphi_1}q_2^{-\lambda}\Big]^{-4}q_2^{3-2\lambda}(q_2^2-q_1^2-q_3^2)(\J^2+8q_1^2q_3^2) \nn\\
&\qquad +\Big[1+\frac{\phi}{\vphi_1}q_3^{-\lambda}\Big]^{-4} q_3^{3-2\lambda}(q_3^2-q_1^2-q_2^2)(\J^2+8q_1^2q_2^2)\Big],
\end{align}
while in cases where $q_3$ may vanish but $q_1$ is nonzero,
\begin{align}
\A(q_1;q_2,q_3) &= \frac{\alpha\pi^2}{6}\frac{1}{\J^4}
\Big[16 q_1^{3}q_2^{3-2\lambda}q_3^{3}\Big[1+\frac{\phi}{\vphi_1}q_2^{-\lambda}\Big]^{-4}  \nn\\&\qquad
+\Big[1+\frac{\phi}{\vphi_1}q_1^{-\lambda}\Big]^{-4}q_1^{3-2\lambda}(q_1^2-q_2^2-q_3^2)(\J^2+8q_2^2q_3^2) \nn\\
&\qquad +\Big[1+\frac{\phi}{\vphi_1}q_2^{-\lambda}\Big]^{-4}q_2^{3-2\lambda}(q_2^2-q_1^2-q_3^2)(\J^2+8q_1^2q_3^2) \nn\\
&\qquad +\Big[1+\frac{\phi}{\vphi_1}q_1^{-\lambda}\Big]^{-4} q_1^{-2\lambda}q_3^{3}(q_3^2-q_1^2-q_2^2)(\J^2+8q_1^2q_2^2)\Big].
\end{align}
In the quasi-equilateral case \eqref{quasi_eq} where all momenta are comparable, both formulae coincide and we obtain simply
\[
\A_{eq}(q_1,q_2,q_3) = \frac{\alpha\pi^2}{6} q_3^{-2\lambda} \Big[1+\frac{\phi}{\vphi_1} q_3^{-\lambda}\Big]^{-4} \frac{(-a_{123}^3+a_{123}b_{123}+c_{123})}{a_{123}^2},
\]
as we could also have found directly using \eqref{A0eq}.  In the main text we only make use of this last result for the quasi-equilateral case.

\section{Ward identities}
\label{app:Ward}

Under a variation of the sources, the variation of the generating functional of connected correlators in the perturbed theory is
\[
\delta W = -\int \d^3x \sqrt{g} \Big(\frac{1}{2}\<T_{ij}\>_s\delta g^{ij}+\<\O\>_s\delta\phi\Big),
\]
where under a dilatation, these sources transform as $\delta_D g^{ij} = -2g^{ij}$ and $\delta_D\phi =-\lambda\phi$.  If the generating functional $W$ is invariant, we then obtain the Ward identity
\[\label{Ward}
\<T(x)\>_s = -\lambda\phi(x)\<\O(x)\>_s,
\]
where the subscript $s$ indicates that this relation holds for any value of the sources.
In general this identity has anomalies reflecting the introduction of a mass scale during renormalisation, however for the correlators we consider with $0<\lambda\ll 1$ these anomalies do not arise.
Starting from the Ward identity \eqref{Ward} and functionally differentiating with respect to the sources before returning them to their background values (namely, $g_{ij}\tto \delta_{ij}$ and $\phi(x) \tto \phi$, for some nonzero constant $\phi$), we obtain nontrivial relations between correlation functions in the perturbed theory.  

After a single functional differentiation, we obtain, e.g.,
\begin{align}
\<T(x_1)\O(x_2)\> &= -\lambda\phi\<\O(x_1)\O(x_2)\>, \\
\<T(x_1)T_{ij}(x_2)\> & = -\lambda\phi\<\O(x_1)T_{ij}(x_2)\>, \\ 
\label{WardTT}
\<T(x_1)T(x_2)\> &= \lambda^2\phi^2\<\O(x_1)\O(x_2)\>,
\end{align}
where we have omitted ultralocal contact terms proportional to $\delta(x_1-x_2)$.  Combining the various relations obtained by performing two functional differentiations, we then find
\begin{align}
\label{Ward_3Ts}
\<T(x_1)T(x_2)T(x_3)\> &=  -\lambda^3\phi^3\<\O(x_1)\O(x_2)\O(x_3)\> 
+\Big[\delta(x_1-x_2)\big(2\<Y^T(x_1)T(x_3)\> \nn\\&\quad
+\lambda^2(\lambda-1)\phi^2\<\O(x_1)\O(x_3)\>\big)   
 + \mathrm{cyclic\,\,perms}\Big], \\[1ex]
\label{Ward_2Ts}
\<T(x_1)T(x_2)T_{ij}(x_3)\> & = \lambda^2\phi^2\<\O(x_1)\O(x_2)T_{ij}(x_3)\>
+\delta(x_1-x_2)\big(2\<Y^T(x_1)T_{ij}(x_3)\>  \nn\\&\quad
+\lambda(1-\lambda)\phi\<\O(x_1)T_{ij}(x_3)\>\big) 
+\Big[\delta(x_1-x_3)\big(2\<T_{ij}(x_1)T(x_3)\>   \nn\\&\quad 
+2\delta^{kl}\<Y^T_{klij}(x_1)T(x_3)\>\big) 
+(x_1\leftrightarrow x_2)\Big],\\[1ex]
\label{Ward_1T}
\<T(x_1)T_{ij}(x_2)T_{kl}(x_3)\> &= -\lambda\phi\<\O(x_1)T_{ij}(x_2)T_{kl}(x_3)\> 
+\delta(x_2-x_3)\big(2\lambda\phi\<\O(x_1)Y^T_{ijkl}(x_2)\> \nn\\&\quad +2\<T(x_1)Y^T_{ijkl}(x_2)\>\big) 
+\Big[\delta(x_1-x_2)\big(2\<T_{ij}(x_1)T_{kl}(x_3)\> \nn\\&\quad +2\delta^{mn}\<Y^T_{mnij}(x_1)T_{kl}(x_3)\>\big)
 + (x_2 \leftrightarrow x_3 \mathrm{\,\,and\,\,} ij\leftrightarrow kl)\Big],
\end{align}
where the local operator $Y^T_{ijkl}$ is defined by taking the variation of the stress tensor with respect to the metric,
see \eqref{YTdef}.
Here we have denoted the trace of this operator as $Y^T \equiv g^{ij}g^{kl}Y^T_{ijkl}$.
In these expressions we have kept all semilocal contact terms but dropped those that are ultralocal, i.e., we have retained contact terms contributing in the limit where two, but not all three, insertions are brought together.

\bibliography{Def1}

\providecommand{\href}[2]{#2}\begingroup\raggedright\begin{thebibliography}{10}

\bibitem{Cheung:2007st}
C.~Cheung, P.~Creminelli, A.~L. Fitzpatrick, J.~Kaplan, and L.~Senatore, {\it
  {The Effective Field Theory of Inflation}},  {\em JHEP} {\bf 0803} (2008)
  014, [\href{http://xxx.lanl.gov/abs/0709.0293}{{\tt arXiv:0709.0293}}].

\bibitem{Antoniadis:2011ib}
I.~Antoniadis, P.~O. Mazur, and E.~Mottola, {\it {Conformal Invariance, Dark
  Energy, and CMB Non-Gaussianity}},  {\em JCAP} {\bf 1209} (2012) 024,
  [\href{http://xxx.lanl.gov/abs/1103.4164}{{\tt arXiv:1103.4164}}].

\bibitem{Maldacena:2011nz}
J.~M. Maldacena and G.~L. Pimentel, {\it {On graviton non-Gaussianities during
  inflation}},  {\em JHEP} {\bf 1109} (2011) 045,
  [\href{http://xxx.lanl.gov/abs/1104.2846}{{\tt arXiv:1104.2846}}].

\bibitem{Hinterbichler:2011qk}
K.~Hinterbichler and J.~Khoury, {\it {The Pseudo-Conformal Universe: Scale
  Invariance from Spontaneous Breaking of Conformal Symmetry}},  {\em JCAP}
  {\bf 1204} (2012) 023, [\href{http://xxx.lanl.gov/abs/1106.1428}{{\tt
  arXiv:1106.1428}}].

\bibitem{Creminelli:2011mw}
P.~Creminelli, {\it {Conformal invariance of scalar perturbations in
  inflation}},  {\em Phys.Rev.} {\bf D85} (2012) 041302,
  [\href{http://xxx.lanl.gov/abs/1108.0874}{{\tt arXiv:1108.0874}}].

\bibitem{Hinterbichler:2012mv}
K.~Hinterbichler, A.~Joyce, and J.~Khoury, {\it {Non-linear Realizations of
  Conformal Symmetry and Effective Field Theory for the Pseudo-Conformal
  Universe}},  {\em JCAP} {\bf 1206} (2012) 043,
  [\href{http://xxx.lanl.gov/abs/1202.6056}{{\tt arXiv:1202.6056}}].

\bibitem{Creminelli:2012ed}
P.~Creminelli, J.~Norena, and M.~Simonovic, {\it {Conformal consistency
  relations for single-field inflation}},  {\em JCAP} {\bf 1207} (2012) 052,
  [\href{http://xxx.lanl.gov/abs/1203.4595}{{\tt arXiv:1203.4595}}].

\bibitem{Hinterbichler:2012nm}
K.~Hinterbichler, L.~Hui, and J.~Khoury, {\it {Conformal Symmetries of
  Adiabatic Modes in Cosmology}},  {\em JCAP} {\bf 1208} (2012) 017,
  [\href{http://xxx.lanl.gov/abs/1203.6351}{{\tt arXiv:1203.6351}}].

\bibitem{Assassi:2012zq}
V.~Assassi, D.~Baumann, and D.~Green, {\it {On Soft Limits of Inflationary
  Correlation Functions}},  \href{http://xxx.lanl.gov/abs/1204.4207}{{\tt
  arXiv:1204.4207}}.

\bibitem{Kehagias:2012pd}
A.~Kehagias and A.~Riotto, {\it {Operator Product Expansion of Inflationary
  Correlators and Conformal Symmetry of de Sitter}},  {\em Nucl.Phys.} {\bf
  B864} (2012) 492--529, [\href{http://xxx.lanl.gov/abs/1205.1523}{{\tt
  arXiv:1205.1523}}].

\bibitem{Kehagias:2012td}
A.~Kehagias and A.~Riotto, {\it {The Four-point Correlator in Multifield
  Inflation, the Operator Product Expansion and the Symmetries of de Sitter}},
  \href{http://xxx.lanl.gov/abs/1210.1918}{{\tt arXiv:1210.1918}}.

\bibitem{Assassi:2012et}
V.~Assassi, D.~Baumann, and D.~Green, {\it {Symmetries and Loops in
  Inflation}},  \href{http://xxx.lanl.gov/abs/1210.7792}{{\tt
  arXiv:1210.7792}}.

\bibitem{McFadden:2009fg}
P.~McFadden and K.~Skenderis, {\it {Holography for Cosmology}},  {\em Phys.
  Rev.} {\bf D81} (2010) 021301, [\href{http://xxx.lanl.gov/abs/0907.5542}{{\tt
  arXiv:0907.5542}}].

\bibitem{McFadden:2010na}
P.~McFadden and K.~Skenderis, {\it {The Holographic Universe}},  {\em J. Phys.
  Conf. Ser.} {\bf 222} (2010) 012007,
  [\href{http://xxx.lanl.gov/abs/1001.2007}{{\tt arXiv:1001.2007}}].

\bibitem{McFadden:2010jw}
P.~McFadden and K.~Skenderis, {\it {Observational signatures of holographic
  models of inflation}},  {\em Proceedings of 12th Marcel Grossmann Meeting on
  General Relativity} (2010) 2315--2323,
  [\href{http://xxx.lanl.gov/abs/1010.0244}{{\tt arXiv:1010.0244}}].

\bibitem{McFadden:2010vh}
P.~McFadden and K.~Skenderis, {\it {Holographic Non-Gaussianity}},  {\em JCAP}
  {\bf 1105} (2011) 013, [\href{http://xxx.lanl.gov/abs/1011.0452}{{\tt
  arXiv:1011.0452}}].

\bibitem{McFadden:2011kk}
P.~McFadden and K.~Skenderis, {\it {Cosmological 3-point correlators from
  holography}},  {\em JCAP} {\bf 1106} (2011) 030,
  [\href{http://xxx.lanl.gov/abs/1104.3894}{{\tt arXiv:1104.3894}}].

\bibitem{Easther:2011wh}
R.~Easther, R.~Flauger, P.~McFadden, and K.~Skenderis, {\it {Constraining
  holographic inflation with WMAP}},  {\em JCAP} {\bf 1109} (2011) 030,
  [\href{http://xxx.lanl.gov/abs/1104.2040}{{\tt arXiv:1104.2040}}].

\bibitem{Bzowski:2011ab}
A.~Bzowski, P.~McFadden, and K.~Skenderis, {\it {Holographic predictions for
  cosmological 3-point functions}},  {\em JHEP} {\bf 1203} (2012) 091,
  [\href{http://xxx.lanl.gov/abs/1112.1967}{{\tt arXiv:1112.1967}}].

\bibitem{Dias:2011in}
M.~Dias, {\it {Cosmology at the boundary of de Sitter using the dS/QFT
  correspondence}},  {\em Phys.Rev.} {\bf D84} (2011) 023512,
  [\href{http://xxx.lanl.gov/abs/1104.0625}{{\tt arXiv:1104.0625}}].

\bibitem{Coriano:2012hd}
C.~Coriano, L.~Delle~Rose, and M.~Serino, {\it {Three and Four Point Functions
  of Stress Energy Tensors in D=3 for the Analysis of Cosmological
  Non-Gaussianities}},  \href{http://xxx.lanl.gov/abs/1210.0136}{{\tt
  arXiv:1210.0136}}.

\bibitem{Strominger:2001gp}
A.~Strominger, {\it {Inflation and the dS / CFT correspondence}},  {\em JHEP}
  {\bf 0111} (2001) 049, [\href{http://xxx.lanl.gov/abs/hep-th/0110087}{{\tt
  hep-th/0110087}}].

\bibitem{Larsen:2002et}
F.~Larsen, J.~P. van~der Schaar, and R.~G. Leigh, {\it {de Sitter holography
  and the cosmic microwave background}},  {\em JHEP} {\bf 04} (2002) 047,
  [\href{http://xxx.lanl.gov/abs/hep-th/0202127}{{\tt hep-th/0202127}}].

\bibitem{Halyo:2002zg}
E.~Halyo, {\it {Holographic inflation}},  {\em JHEP} {\bf 0402} (2004) 062,
  [\href{http://xxx.lanl.gov/abs/hep-th/0203235}{{\tt hep-th/0203235}}].

\bibitem{Larsen:2003pf}
F.~Larsen and R.~McNees, {\it {Inflation and de Sitter holography}},  {\em
  JHEP} {\bf 07} (2003) 051,
  [\href{http://xxx.lanl.gov/abs/hep-th/0307026}{{\tt hep-th/0307026}}].

\bibitem{vanderSchaar:2003sz}
J.~P. van~der Schaar, {\it {Inflationary perturbations from deformed CFT}},
  {\em JHEP} {\bf 01} (2004) 070,
  [\href{http://xxx.lanl.gov/abs/hep-th/0307271}{{\tt hep-th/0307271}}].

\bibitem{Larsen:2004kf}
F.~Larsen and R.~McNees, {\it {Holography, diffeomorphisms, and scaling
  violations in the CMB}},  {\em JHEP} {\bf 07} (2004) 062,
  [\href{http://xxx.lanl.gov/abs/hep-th/0402050}{{\tt hep-th/0402050}}].

\bibitem{Seery:2006tq}
D.~Seery and J.~E. Lidsey, {\it {Non-Gaussian inflationary perturbations from
  the dS/CFT correspondence}},  {\em JCAP} {\bf 0606} (2006) 001,
  [\href{http://xxx.lanl.gov/abs/astro-ph/0604209}{{\tt astro-ph/0604209}}].

\bibitem{Strominger:2001pn}
A.~Strominger, {\it {The dS/CFT correspondence}},  {\em JHEP} {\bf 10} (2001)
  034, [\href{http://xxx.lanl.gov/abs/hep-th/0106113}{{\tt hep-th/0106113}}].

\bibitem{Witten:2001kn}
E.~Witten, {\it {Quantum gravity in de Sitter space}},
  \href{http://xxx.lanl.gov/abs/hep-th/0106109}{{\tt hep-th/0106109}}.

\bibitem{Maldacena:2002vr}
J.~M. Maldacena, {\it {Non-Gaussian features of primordial fluctuations in
  single field inflationary models}},  {\em JHEP} {\bf 05} (2003) 013,
  [\href{http://xxx.lanl.gov/abs/astro-ph/0210603}{{\tt astro-ph/0210603}}].

\bibitem{Harlow:2011ke}
D.~Harlow and D.~Stanford, {\it {Operator Dictionaries and Wave Functions in
  AdS/CFT and dS/CFT}},  \href{http://xxx.lanl.gov/abs/1104.2621}{{\tt
  arXiv:1104.2621}}.

\bibitem{Dong:2011uf}
X.~Dong, B.~Horn, S.~Matsuura, E.~Silverstein, and G.~Torroba, {\it {FRW
  solutions and holography from uplifted AdS/CFT}},  {\em Phys.Rev.} {\bf D85}
  (2012) 104035, [\href{http://xxx.lanl.gov/abs/1108.5732}{{\tt
  arXiv:1108.5732}}].

\bibitem{Anninos:2011ui}
D.~Anninos, T.~Hartman, and A.~Strominger, {\it {Higher Spin Realization of the
  dS/CFT Correspondence}},  \href{http://xxx.lanl.gov/abs/1108.5735}{{\tt
  arXiv:1108.5735}}.

\bibitem{Anninos:2011af}
D.~Anninos, S.~A. Hartnoll, and D.~M. Hofman, {\it {Static Patch Solipsism:
  Conformal Symmetry of the de Sitter Worldline}},  {\em Class.Quant.Grav.}
  {\bf 29} (2012) 075002, [\href{http://xxx.lanl.gov/abs/1109.4942}{{\tt
  arXiv:1109.4942}}].

\bibitem{Hertog:2011ky}
T.~Hertog and J.~Hartle, {\it {Holographic No-Boundary Measure}},  {\em JHEP}
  {\bf 1205} (2012) 095, [\href{http://xxx.lanl.gov/abs/1111.6090}{{\tt
  arXiv:1111.6090}}].

\bibitem{Hartle:2012qb}
J.~B. Hartle, S.~Hawking, and T.~Hertog, {\it {Accelerated Expansion from
  Negative $\Lambda$}},  \href{http://xxx.lanl.gov/abs/1205.3807}{{\tt
  arXiv:1205.3807}}.

\bibitem{Anninos:2012ft}
D.~Anninos, F.~Denef, and D.~Harlow, {\it {The Wave Function of Vasiliev's
  Universe - A Few Slices Thereof}},
  \href{http://xxx.lanl.gov/abs/1207.5517}{{\tt arXiv:1207.5517}}.

\bibitem{Hartle:2012tv}
J.~B. Hartle, S.~Hawking, and T.~Hertog, {\it {Inflation with Negative
  $\Lambda$}},  \href{http://xxx.lanl.gov/abs/1207.6653}{{\tt
  arXiv:1207.6653}}.

\bibitem{Castro:2012gc}
A.~Castro and A.~Maloney, {\it {The Wave Function of Quantum de Sitter}},
  \href{http://xxx.lanl.gov/abs/1209.5757}{{\tt arXiv:1209.5757}}.

\bibitem{Marolf:2012kh}
D.~Marolf, I.~A. Morrison, and M.~Srednicki, {\it {Perturbative S-matrix for
  massive scalar fields in global de Sitter space}},
  \href{http://xxx.lanl.gov/abs/1209.6039}{{\tt arXiv:1209.6039}}.

\bibitem{Ludwig:1987gs}
A.~Ludwig and J.~L. Cardy, {\it {Perturbative Evaluation of the Conformal
  Anomaly at New Critical Points with Applications to Random Systems}},  {\em
  Nucl.Phys.} {\bf B285} (1987) 687--718.

\bibitem{Zamolodchikov:1987ti}
A.~Zamolodchikov, {\it {Renormalization Group and Perturbation Theory Near
  Fixed Points in Two-Dimensional Field Theory}},  {\em Sov.J.Nucl.Phys.} {\bf
  46} (1987) 1090.

\bibitem{Komatsu:2010fb}
E.~Komatsu {\em et.~al.}, {\it {Seven-Year Wilkinson Microwave Anisotropy Probe
  (WMAP) Observations: Cosmological Interpretation}},
  \href{http://xxx.lanl.gov/abs/1001.4538}{{\tt arXiv:1001.4538}}.

\bibitem{Schalm:2012pi}
K.~Schalm, G.~Shiu, and T.~van~der Aalst, {\it {Consistency condition for
  inflation from (broken) conformal symmetry}},
  \href{http://xxx.lanl.gov/abs/1211.2157}{{\tt arXiv:1211.2157}}.

\bibitem{Polyakov:1987ez}
A.~M. Polyakov, {\it {Gauge fields and strings}},  {\em Contemp. Concepts
  Phys.} {\bf 3} (1987) 1--301.

\bibitem{Klebanov:2011gs}
I.~R. Klebanov, S.~S. Pufu, and B.~R. Safdi, {\it {F-Theorem without
  Supersymmetry}},  {\em JHEP} {\bf 1110} (2011) 038,
  [\href{http://xxx.lanl.gov/abs/1105.4598}{{\tt arXiv:1105.4598}}].

\bibitem{Osborn:1993cr}
H.~Osborn and A.~Petkou, {\it {Implications of conformal invariance in field
  theories for general dimensions}},  {\em Annals Phys.} {\bf 231} (1994)
  311--362, [\href{http://xxx.lanl.gov/abs/hep-th/9307010}{{\tt
  hep-th/9307010}}].

\bibitem{Cardy:1987dg}
J.~L. Cardy, {\it {Anisotropic corrections to correlation functions in finite
  size systems}},  {\em Nucl.~Phys.} {\bf B290} (1987) 355--362.

\bibitem{Cvetic:1994ya}
M.~Cvetic and H.~H. Soleng, {\it {Naked singularities in dilatonic domain wall
  space times}},  {\em Phys. Rev.} {\bf D51} (1995) 5768--5784,
  [\href{http://xxx.lanl.gov/abs/hep-th/9411170}{{\tt hep-th/9411170}}].

\bibitem{Skenderis:2006jq}
K.~Skenderis and P.~K. Townsend, {\it {Hidden supersymmetry of domain walls and
  cosmologies}},  {\em Phys. Rev. Lett.} {\bf 96} (2006) 191301,
  [\href{http://xxx.lanl.gov/abs/hep-th/0602260}{{\tt hep-th/0602260}}].

\bibitem{Salopek:1990jq}
D.~S. Salopek and J.~R. Bond, {\it {Nonlinear evolution of long wavelength
  metric fluctuations in inflationary models}},  {\em Phys. Rev.} {\bf D42}
  (1990) 3936--3962.

\bibitem{Freedman:2003ax}
D.~Z. Freedman, C.~Nunez, M.~Schnabl, and K.~Skenderis, {\it {Fake Supergravity
  and Domain Wall Stability}},  {\em Phys. Rev.} {\bf D69} (2004) 104027,
  [\href{http://xxx.lanl.gov/abs/hep-th/0312055}{{\tt hep-th/0312055}}].

\bibitem{Skenderis:2002wp}
K.~Skenderis, {\it {Lecture notes on holographic renormalization}},  {\em
  Class. Quant. Grav.} {\bf 19} (2002) 5849--5876,
  [\href{http://xxx.lanl.gov/abs/hep-th/0209067}{{\tt hep-th/0209067}}].

\bibitem{Stewart:1993bc}
E.~D. Stewart and D.~H. Lyth, {\it {A More accurate analytic calculation of the
  spectrum of cosmological perturbations produced during inflation}},  {\em
  Phys.Lett.} {\bf B302} (1993) 171--175,
  [\href{http://xxx.lanl.gov/abs/gr-qc/9302019}{{\tt gr-qc/9302019}}].

\bibitem{Gong:2001he}
J.-O. Gong and E.~D. Stewart, {\it {The density perturbation power spectrum to
  second order corrections in the slow roll expansion}},  {\em Phys.Lett.} {\bf
  B510} (2001) 1--9, [\href{http://xxx.lanl.gov/abs/astro-ph/0101225}{{\tt
  astro-ph/0101225}}].

\bibitem{Freedman:1991tk}
D.~Z. Freedman, K.~Johnson, and J.~I. Latorre, {\it {Differential
  regularization and renormalization: A New method of calculation in quantum
  field theory}},  {\em Nucl.Phys.} {\bf B371} (1992) 353--414.

\bibitem{Prudnikov}
Y.~Brychkov, A.~Prudnikov, and O.~Marichev, {\em Tables of Indefinite
  Integrals, vol.~2}.
\newblock Gordon \& Breach Science, 1989.

\bibitem{Boos:1987bg}
E.~Boos and A.~I. Davydychev, {\it {A method of evaluation of vertex type
  Feynman integrals}},  {\em Moscow Univ.Phys.Bull.} {\bf 42N3} (1987) 6--10.

\bibitem{Anastasiou:1999ui}
C.~Anastasiou, E.~N. Glover, and C.~Oleari, {\it {Scalar one loop integrals
  using the negative dimension approach}},  {\em Nucl.Phys.} {\bf B572} (2000)
  307--360, [\href{http://xxx.lanl.gov/abs/hep-ph/9907494}{{\tt
  hep-ph/9907494}}].

\bibitem{Exton}
H.~Exton, {\it {On the system of partial differential equations associated with
  Appell's function $F_4$}},  {\em J.~Phys.~A:~Math.~Gen.} {\bf 28} (1995)
  631--641.

\bibitem{Alkofer:2008dt}
R.~Alkofer, M.~Q. Huber, and K.~Schwenzer, {\it {Infrared Behavior of
  Three-Point Functions in Landau Gauge Yang-Mills Theory}},  {\em Eur.Phys.J.}
  {\bf C62} (2009) 761--781, [\href{http://xxx.lanl.gov/abs/0812.4045}{{\tt
  arXiv:0812.4045}}].

\bibitem{Bateman}
A.~Erd{\'e}lyi~(editor), {\em Bateman manuscript project, Higher transcendental
  functions, vol.~1, p.~238}.
\newblock McGraw-Hill, 1953.

\end{thebibliography}\endgroup
\bibliographystyle{jhep} 

\end{document}